%% file: main.tex
\documentclass[twocolumn,aps,prb,superscriptaddress,amsfonts,longbibliography]{revtex4-1}
\usepackage[tight]{subfigure}
\usepackage{physics,tensor,siunitx}
\usepackage{amsmath,dsfont}
\usepackage{mathtools}
\usepackage{amssymb}
\usepackage{booktabs}
\usepackage{mathrsfs}
\usepackage{varwidth}
\usepackage{graphicx,color,caption}
\usepackage{bm}
\usepackage{cases}
\usepackage{array}
\usepackage{dcolumn}
\usepackage{chngcntr}
\usepackage{hepparticles}
\usepackage{heppennames}
\usepackage{hepnicenames}
\usepackage{epstopdf}
\usepackage{newfloat}
\usepackage{nicefrac}
\usepackage{comment}
\usepackage{svg}
\usepackage{etex}

\DeclareFloatingEnvironment[name={Extended Data Figure}]{suppfigure}
\DeclareUnicodeCharacter{2212}{-}
\usepackage{hyperref}
\hypersetup{
    colorlinks=true,
    linkcolor=blue,     
    citecolor=blue,    
    urlcolor={}         
}
\usepackage[capitalise]{cleveref}
\usepackage{tikz}
\usetikzlibrary{decorations.pathmorphing} 
\usetikzlibrary{fit}					
\usetikzlibrary{backgrounds}	
\usetikzlibrary{shapes.geometric} %
\usetikzlibrary{positioning}
\usetikzlibrary{arrows.meta}
\usetikzlibrary{quantikz}
\usepackage{yquant}

\begin{document}

\title{Hidden Prime-Factor Subgroups in Molecular and Condensed-Phase Systems}

\author{Srinivasan S. Iyengar}
\email{iyengar@iu.edu}
\affiliation{Department of Chemistry and Department of Physics, Indiana University, Bloomington, Indiana 47405, USA}
\affiliation{Indiana University Quantum Science and Engineering Center, Bloomington, Indiana 47405, USA}

\author{Amr Sabry}
\affiliation{Department of Computer Science, Luddy School of Informatics, Computing, and Engineering, Indiana University}
\affiliation{Indiana University Quantum Science and Engineering Center, Bloomington, Indiana 47405, USA}

\date{\today}

\begin{abstract}
We describe a group theoretic analysis of Shor's algorithm and other related hidden subgroup problems in mathematics and relate these to symmetries of molecular and condensed phase assemblies. By recasting Shor's algorithm through the lens of group theory, we expose the possibility that physical systems such as molecular orbitals, condensed phase assemblies and optical beams may be designed such that these contain information pertaining to the solution to hard mathematical problems such as prime-factoring. We discuss real molecular systems, whose orbitals are constructed from symmetry-adapted linear combinations of atomic orbitals, and show that these contain information pertaining to the prime-factors of corresponding integers. Due to the broad significance of prime-factoring towards a variety of encryption problems in cyber-security, we believe this novel and fundamental approach may have broad impact. 
\end{abstract}
\maketitle

\section{Introduction}

Group theory has long provided a unifying language for understanding structure and dynamics in molecular, optical, and condensed-phase systems\cite{Tinkham,hamermesh,cotton}. In molecular spectroscopy\cite{cotton,ewk} and in the analysis of chemical transformations\cite{Tinkham,Albright2013}, symmetry constrains allowed transitions, organizes energy levels, and determines the form of wavefunctions through irreducible representations and symmetry-adapted linear combinations. These same representation-theoretic tools appear throughout quantum physics, from Bloch’s theorem in periodic media to the construction of molecular orbitals and band structures.

Independently, modern number theory and quantum algorithms have shown that arithmetic problems such as order-finding and integer factorization admit natural group-theoretic formulations. Modular arithmetic over $\mathbb{Z}_N$ forms a cyclic group, whose decomposition into prime-order subgroups, formalized by the Chinese Remainder Theorem (CRT)\cite{ChineseRemainderTheorem}, governs the periodic structure underlying Shor’s algorithm\cite{Shor}. In the standard hidden-subgroup formulation\cite{simon,deutsch,deutschJozsa,Shor,Vazirani-Bernstein,shorFermat,Sabry-Iyengar-2023}, factoring reduces to extracting the period of a function over an extended cyclic group, a task performed by Fourier analysis\cite{Nielsen-Chuang-QuantComp}. 

Although these two uses of symmetry, physical and arithmetic, have traditionally been studied in separate domains, they rely on the same underlying representation-theoretic machinery. Cyclic groups, irreducible representations, Fourier modes, and projection operators appear both in the analysis of molecular wavefunctions and in the structure of quantum algorithms for factoring. This observation motivates the central question of this work: can physical systems whose wavefunctions are organized by cyclic symmetries, or optical beams with precise angular momenta, serve as faithful realizations of the group-theoretic structure that underlies arithmetic computation? 

This general question also relates to our previous efforts\cite{Sabry-Iyengar-2023} where we discuss the similarities between several textbook quantum algorithms, including the Deutsch\cite{deutsch}, Deutsch-Jozsa\cite{deutschJozsa}, Bernstein-Vazirani\cite{Vazirani-Bernstein}, Simon\cite{simon}, and Shor\cite{Shor} algorithms, and natural quantum phenomena described using conventional approaches to quantum dynamics. These textbook quantum algorithms are thought to be exponentially faster than their best known classical counterparts, and are algorithms considered to be solving instances of the {hidden subgroup problem}.  Interestingly, all of these algorithms have a similar mathematical structure, as shown in Figure \ref{fig:templateQC}, and discussed in detailed in Ref. \onlinecite{Sabry-Iyengar-2023}.
\begin{figure}[t]
  \begin{tikzpicture}[scale=0.7,every label/.style={rotate=40, anchor=south west}]
    \begin{yquant*}[operators/every barrier/.append style={red, thick},
        operator/minimum width=7mm,
        operator/separation=1mm,
        register/separation=10mm]
    qubits {$\ket0^{\otimes n}$} a;
    qubits {$\ket{\psi}_m$} b;
    box {$H^{\otimes n}$} a;
    barrier (-);
    [x radius=7mm, y radius=7mm]
    box {$U_f$} (a,b);
    barrier (-);
    measure b;
    discard b;
    barrier (-);
    box {$\mathit{QFT}$} a;
    measure a;
    \end{yquant*}
  \end{tikzpicture}
\caption{\label{fig:templateQC}A common template quantum circuit for all hidden subgroup problems\cite{Sabry-Iyengar-2023}. All these algorithms begin by creating an equal superposition of all relevant possibilities (the Hadamard gates on the ancilla above), apply the $U_f$ block to the superposition, and analyze the result using the Quantum Fourier Transform (QFT) or a similar rotation into a complementary space.}
\end{figure}
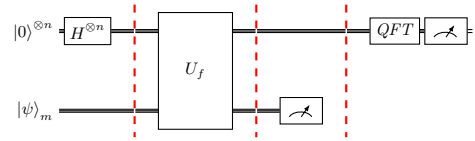

\begin{figure*}[t]
\centering
\begin{tikzpicture}[
    every node/.style={font=\small},
    shor/.style={rectangle, rounded corners=3pt, fill=purple!15,
                 draw=purple!60, line width=0.6pt,
                 text width=3.8cm, align=center,
                 minimum height=1.0cm, inner sep=4pt},
    mol/.style={rectangle, rounded corners=3pt, fill=teal!15,
                draw=teal!60, line width=0.6pt,
                text width=3.8cm, align=center,
                minimum height=1.0cm, inner sep=4pt},
    bloch/.style={rectangle, rounded corners=3pt, fill=orange!12,
                draw=orange!60, line width=0.6pt,
                text width=3.8cm, align=center,
                minimum height=1.0cm, inner sep=4pt},
    shared/.style={rectangle, rounded corners=3pt, fill=gray!10,
                   draw=gray!50, line width=0.6pt,
                   text width=2.0cm, align=center,
                   minimum height=1.0cm, inner sep=4pt},
    tentmol/.style={rectangle, rounded corners=3pt, fill=white,
                draw=teal!90, line width=2.0pt,
                text width=3.8cm, align=center,
                minimum height=1.0cm, inner sep=4pt, dashed},
    tentbloch/.style={rectangle, rounded corners=3pt, fill=white,
                draw=orange!90, line width=2.0pt,
                text width=3.8cm, align=center,
                minimum height=1.0cm, inner sep=4pt, dashed},
    arr/.style={->, >=stealth, thick, gray!70},
    harr/.style={->, >=stealth, thin, gray!60},
]

\def\xS{-6.11}
\def\xSh{-2.34}
\def\xM{1.43}
\def\xB{6.10}
\def\dy{3.0}   

\node[shor,  fill=purple!35, draw=purple!70, line width=1pt,
      minimum height=0.8cm, font=\small\bfseries]
    at (\xS,  0) {Shor's algorithm};
\node[shared, fill=gray!22, draw=gray!60, line width=1pt,
      minimum height=0.8cm, font=\small\bfseries]
    at (\xSh, 0) {Shared structure};
\node[mol,   fill=teal!30,   draw=teal!70,   line width=1pt,
      minimum height=0.8cm, font=\small\bfseries]
    at (\xM,  0) {Molecular orbitals};
\node[bloch, fill=orange!28, draw=orange!70, line width=1pt,
      minimum height=0.8cm, font=\small\bfseries]
    at (\xB,  0) {Bloch wavefunctions};

\node[shor] (s1) at (\xS, -1*\dy) {
    \textbf{Extended cyclic group}\\[1pt]
    Construct $\mathcal{G}^{N,a}$ from $N$ and coprime $a$};
\node[shared] (m1) at (\xSh, -1*\dy) {
    $\mathbb{Z}_N$ as finite rotation group};
\node[mol] (t1) at (\xM, -1*\dy) {
    \textbf{Molecular symmetry}\\[1pt]
    $\mathcal{G}^{N}$ from $N$-fold point group};
\node[bloch] (b1) at (\xB, -1*\dy) {
    \textbf{Lattice wavefunctions}\\[1pt]
    Periodic potential period $2\pi/N$: Translation group
    };

\node[shor] (s2) at (\xS, -2*\dy) {
    \textbf{Apply oracle}\\[1pt]
    $\sum_x e^{i(2\pi/N)jx}\, a^x \bmod N$};
\node[shared] (m2) at (\xSh, -2*\dy) {
    Phase-weighted superposition of translated functions};
\node[mol] (t2) at (\xM, -2*\dy) {
    \textbf{Projection operators}\\[1pt]
    $\mathcal{P}^{(j)}f_{\rm AO}
     =\frac{1}{N}\!\sum_k e^{i\frac{2\pi}{N}jk}
       f_{\rm AO}(C_N^{N-k}x)$};
\node[bloch] (b2) at (\xB, -2*\dy) {
    \textbf{Bloch 
    decomposition}\\[1pt]
    $\psi_k(x)=e^{ikx}u_k(x)$
    };

\node[shor] (s3) at (\xS, -3*\dy) {
    \textbf{Coset decomposition}\\[1pt]
    $f(x)$ constant on cosets of $\mathcal{G}^{a}$; $\mathcal{G}^N$ hidden};
\node[shared] (m3) at (\xSh, -3*\dy) {
    Chinese Remainder Theorem};
\node[mol] (t3) at (\xM, -3*\dy) {
    \textbf{Prime-factor sub-groups 
    }\\[1pt]
    SALCs factor as $\bigotimes_i \mathcal{P}^{(j_i)}_{b_i} f_{\rm AO}$};
\node[bloch] (b3) at (\xB, -3*\dy) {
    \textbf{Prime-factor sub-group band structure}\\[1pt]
    Bloch phases $e^{ik(2\pi/N)}$ decompose through
    prime-order momenta $k_{b_i}$};

\node[shor] (s4) at (\xS, -4*\dy) {
    \textbf{Quantum Fourier transform}\\[1pt]
    Peaks at multiples of $aN/r$;\linebreak yields period $r$};
\node[shared] (m4) at (\xSh, -4*\dy) {
    Fourier analysis extracts period};
\node[tentmol] (t4) at (\xM, -4*\dy) {
    \textbf{Spectral readout}$\,^\dagger$\\[1pt]
    Fourier modes of molecular wavefunctions labeled by prime-factor subgroup index $b_i$};
\node[tentbloch] (b4) at (\xB, -4*\dy) {
    \textbf{Spectral readout}$\,^\dagger$\\[1pt]
    Fourier modes of solids labeled by prime-factor subgroup index $b_i$};

\node[shor] (s5) at (\xS, -5*\dy) {
    \textbf{Prime factors of $N$}\\[1pt]
    via $\gcd(a^{r/2}\!\pm\!1,\,N)$};
\node[shared] (m5) at (\xSh, -5*\dy) {
    Same arithmetic answer};
\node[tentmol] (t5) at (\xM, -5*\dy) {
    \textbf{Prime factors of $N$}$\,^\dagger$\\[1pt]
    encoded in orbital Fourier structure};
\node[tentbloch] (b5) at (\xB, -5*\dy) {
    \textbf{Prime factors of $N$}$\,^\dagger$\\[1pt]
    encoded in Bloch states;\linebreak
    quantum corrals as engineered analogues};

\node[draw=gray!50, line width=1.2pt, dotted, rounded corners=5pt,
      fit=(t1)(t2)(t3)(t4)(t5)(b1)(b2)(b3)(b4)(b5),
      inner sep=6pt, label={[gray!70, font=\small\bfseries]above:Physical realizations}] {};
      
\foreach \i/\j in {1/2, 2/3, 3/4, 4/5} {
    \draw[arr] (s\i) -- (s\j);
    \draw[arr] (t\i) -- (t\j);
    \draw[arr] (b\i) -- (b\j);
}

\node[font=\footnotesize, text width=15.5cm, align=left]
    at (-0.01, -5*\dy - 1.8)
    {$^\dagger$~Established here as a structural correspondence;
     experimental realization in engineered molecular, condensed-phase,
     and optical systems remains an open question.};

\end{tikzpicture}
\caption{Step-by-step correspondence between Shor's algorithm (left),
the construction of symmetry-adapted molecular orbitals (center-right),
and Bloch-wave / optical systems (far right).
The shared mathematical structure at each step (center) is an operational identity.
Speculative steps---whose experimental realization remains open---are shown
with pale fills and dashed borders ($^\dagger$).}
\label{fig:correspondence}
\end{figure*}

Here we use symmetry to probe connections between arithmetic problems and physical systems and ask if we may be able to design molecular or condensed phase  systems that have properties related to solutions to hidden subgroup problems. In Figure~\ref{fig:correspondence} we summarize the step-by-step 
correspondence established in this paper. Specifically, at each stage of 
Shor's algorithm there is an operationally identical step in 
the construction of symmetry-adapted molecular orbitals, with 
the Chinese Remainder Theorem playing the same structural role 
in both. 

Accordingly, we proceed as follows. In Section~\ref{sec:one}, we introduce modular arithmetic in group-theoretic form by representing $\mathbb{Z}_N$ as a finite rotation group. This makes its cyclic structure geometrically explicit and allows the prime-factor decomposition of $N$ to be expressed as a decomposition of the group into prime-order cyclic subgroups via the Chinese Remainder Theorem. Section~\ref{sec:two} develops the representation theory of these cyclic groups. We introduce irreducible representations and projection operators and show how symmetry-adapted functions on $\mathbb{Z}_N$ factor naturally along the prime-order subgroups determined by the factorization of $N$. In Section~\ref{sec:three}, we turn to the group-theoretic formulation of Shor’s algorithm. We construct the extended cyclic group labeled as ${\cal G}^{N,a}$ generated by $N$ and a coprime element $a$, and we show how the periodicity of the function $a^x \bmod N$ corresponds to hidden cyclic subgroups whose structure reflects the prime factors of $N$. Section~\ref{sec:four} reformulates this hidden-subgroup structure in purely representation-theoretic terms, making explicit the role of Fourier modes and coset decompositions in extracting arithmetic periodicity. In Section~\ref{sec:five}, we connect these abstract constructions to molecular and condensed phase systems  via Bloch’s theorem, showing that wavefunctions in periodic systems realize the same cyclic symmetry and Fourier structure as functions on $\mathbb{Z}_N$. Section~\ref{sec:five} also introduces symmetry-adapted linear combinations (SALCs)\cite{cotton,Tinkham} as the physical implementation of projection operators, demonstrating how molecular orbitals and condensed-matter states realize irreducible representations of cyclic groups and their prime-order subgroups. 

Thus, 
we discuss how molecular, and condensed-phase, systems with engineered cyclic symmetries can act as physical analogues of the group-theoretic structures underlying factoring, and we outline how their spectral and dynamical properties may be used to access arithmetic periodicity. Through this progression, we show that the same representation-theoretic framework governs modular arithmetic, Shor’s algorithm, and symmetry-adapted wavefunctions of molecules, opening a path toward viewing physical symmetry not merely as an organizing principle for matter, but as a computational resource for arithmetic structure.

\section{Cyclic Groups and Modular Arithmetic}
\label{sec:one}

\subsection{Modular Arithmetic as a Cyclic Group}

Consider the number $N$ with prime-factors given by $\left\{ b_i \right\}$, and corresponding multiples, $\left\{ m_i \right\}$. That is, $m_ib_i=N$, for all $\left\{ m_i, b_i \right\}$. We introduce a rotation group, that is, a finite subgroup of $SO(2)$, as follows. 
Consider 
the rotation operation, $R_z(2\pi/N)$, 
as a representation for the noted rotation angle about the $z$-axis in three-dimensional Euclidean space. We represent the operation $R_z(2\pi/N) \equiv C_{N}$ and 
we will treat this as a fundamental rotation and maps to the integer ``$1$'', 
and associated with such operations, we introduce a group with elements given by
\begin{align}
    {\cal G}^N = \left\{ C_{N}, C_{N}^2, \cdots, C_{N}^{b_1}, \cdots, C_{N}^{b_2}, \cdots, C_{N}^{N-1}, E \right\}
    \label{GN-group}
\end{align}
Here $C_{N}^j\equiv R_z(2\pi j/N)$, $E=C_{N}^{N}$, is the identity operation, and we have individually identified the operations $C_N^{b_i} \equiv {C_{N}}\cdot{C_{N}}\cdot{C_{N}}\cdots b_i \text{times}$, within the group to show that these rotations involving the prime factors are also members of the group, ${\cal G}^N$. Thus each rotation operation above represents a single integer in the domain $[0,N]$. Additionally, in standard notations,\cite{Tinkham,cotton} the group as well as rotation operations are represented as $C_N$, but we have changed our convention above to maintain clarity and for reasons that will become clear later in the paper. 

It is trivial to show that Eq. (\ref{GN-group}) forms a group\cite{Tinkham} because the group is closed under rotations, all elements contain an inverse, and the identity operation is also present within the group. Additionally, the group is associative. 
Thus, we can visualize this as the unit circle being divided into pizza slices such that each incremental rotation angle corresponds to one slice and one integer number and the circle is divided into $N$ equal angle slices. In that case, the rotation $C_{N}^{b_i}$ corresponds to the $b_i$-th slice and corresponds to the integer $b_i$. In a similar manner, the set of integers $[0,\cdots,N]$, also forms a group under addition with modulo remainder operations. In this sense, the group ${\cal G}^N$ is isomorphic to the group created by a natural number $N$ and represented as $\mathbb{Z}_N$ in  mathematics\cite{Burton2011}, that is ${\cal G}^N \equiv \mathbb{Z}_N$.

\begin{figure}
    \centering
    \subfigure[]{\includegraphics[width=0.4\linewidth]{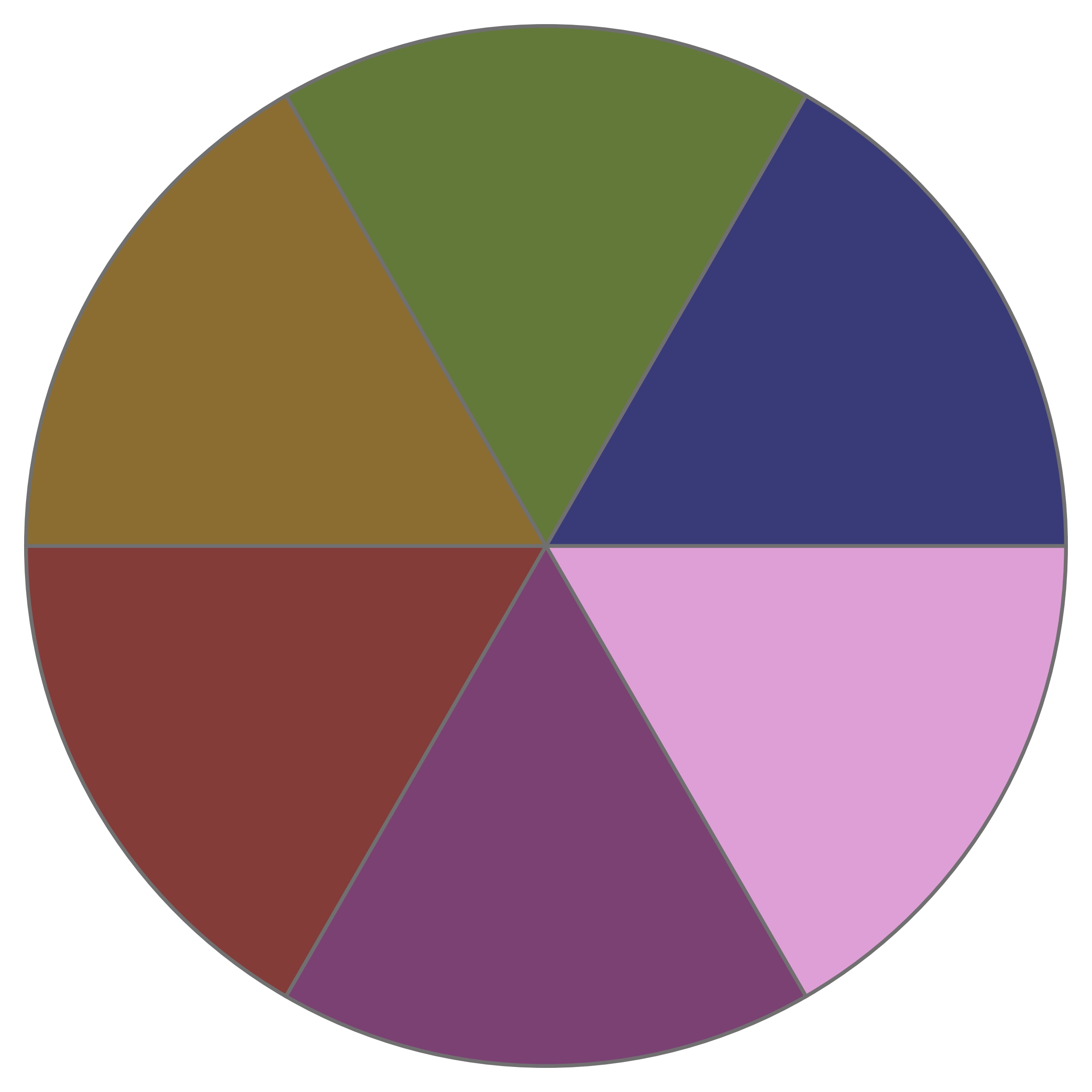}}
    \subfigure[]{\includegraphics[width=0.4\linewidth]{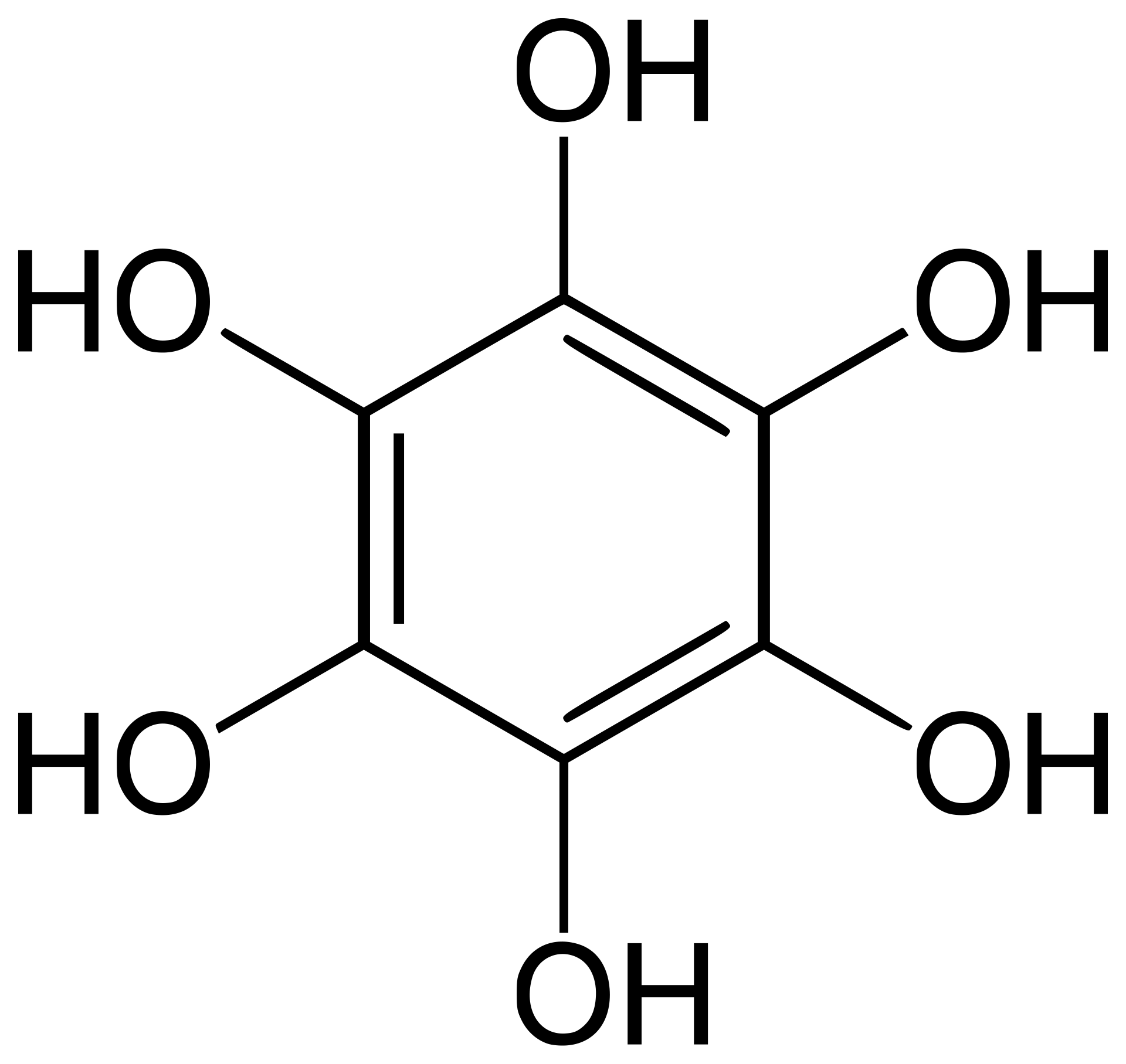}}      \caption{Illustration  of (a): ${\cal G}^{6}$, The circle is divided into $N$ equal portions and the colors are chosen purely to make the regions distinct. The isomorphic objects being constructed here are symmetric with rotations, (Figure (a) was generated using the AI tool Microsoft Copilot.) Figure (b) is the molecule hexabydroxybenzene (or benzenehexol), with the OH groups pointing out of the plane, that has the same symmetry as ${\cal G}^{6}$.}
    \label{fig:G-6-1}
\end{figure}
To illustrate Eq. (\ref{GN-group}), let us consider $N=6$, with prime-factors $b_1=2$ and $b_2=3$. The associated group is
\begin{align}
    {\cal G}^6 = \left\{ C_{6}, C_{6}^2, C_{6}^{3}, C_{6}^{4},  C_{6}^{5}, E \right\}
\end{align}
The molecule, hexahydroxybenzene, is an example of a system that obeys such a symmetry, and the molecular orbitals of hexahydroxybenzene capture this symetry. We can readily see that $C_{N}^{b_i}$ are part of the group, that is $C_{6}^2, C_{6}^{3}$, and their inverse operations, $ C_{6}^{4},  C_{6}^{3}$ are also part of the group. The group ${\cal G}^6$ and the molecules hexahydroxybenzene are illustrated in Figure \ref{fig:G-6-1}. 

\subsection{Prime-Order Subgroups and the Chinese Remainder Theorem}

We now consider the groups ${\cal G}^{b_i}$, that are subgroups of ${\cal G}^{N}$. Given that the $b_i$ are prime numbers, these groups contain 
\begin{align}
    {\cal G}^{b_i} = \left\{ C_{b_i}, C_{b_i}^2, \cdots, E \right\}
    \label{Gbi-group}
\end{align}
where $C_{b_i}=R_z(2\pi/{b_i})=R_z(m_i 2\pi/{N})=C_N^{m_i}$, and as before, is the fundamental rotation and maps on to the integer ``$1$'' in ${\cal G}^{b_i}$. According to the Chinese Remainder Theorem (CRT)\cite{ChineseRemainderTheorem}, the group ${\cal G}^{N}$ can be created from tuples (direct products) of $\left\{ {\cal G}^{b_i} \right\}$. That is, 
\begin{align}
    {\cal G}^{N} \equiv \left\{ {\cal G}^{b_i} {\cal G}^{b_j}\; \vert \; \forall \; b_i, b_j \right\}
    \label{Gbi-group}
\end{align}
Let us illustrate these concepts for the case of benzene, that is $N=6$. Here the groups ${\cal G}^{b_i}$ are given by,
\begin{align}
    {\cal G}^{2} = \left\{ C_{2}, E \right\} \equiv \left\{ C_6^3, C_6^6 \right\}
    \label{G2-group}
\end{align}
\begin{align}
    {\cal G}^{3} = \left\{ C_{3}, C_3^2, E \right\} \equiv \left\{ C_6^2, C_6^4, C_6^6 \right\}
    \label{G3-group}
\end{align}
and correspondingly
\begin{align}
    {\cal G}^6 =& \left\{ {\cal G}^{2} {\cal G}^{3} \right\} = {\cal G}^{2} \otimes {\cal G}^{3} \nonumber \\ 
    =& \left\{ C_6^3, C_6^6 \equiv E, \right. \nonumber \\ & \left. \phantom{[} C_6^2, C_6^4, C_6^6 \equiv E, \right. \nonumber \\ & \left. \phantom{[} C_6^3 C_{6}^2 \equiv C_6^5, C_6^3 C_6^4 \equiv C_6, \right\} \nonumber \\ 
    =& \left\{ C_{6}, C_{6}^2, C_{6}^{3}, C_{6}^{4},  C_{6}^{5}, E \right\}
    \label{CRT-6}
\end{align}
Likewise, for $N=15$ we have, 
\begin{align}
    {\cal G}^{15} = \left\{ C_{15}, C_{15}^2, C_{15}^3, C_{15}^4, C_{15}^5, \cdots   E \right\}
\end{align}
and for $N=21$, 
\begin{align}
    {\cal G}^{21} = \left\{ C_{21}, C_{21}^2, C_{21}^3, \cdots, C_{21}^7, \cdots  E \right\}
\end{align}
However we can also write:
\begin{align}
    {\cal G}^{15} =& \left\{ {\cal G}^{3} {\cal G}^{5} \right\} 
    \label{CRT-15-1}
\end{align}
It is trivial to show that the two representations of ${\cal G}^{15}$ above are equivalent. For example, $C_{15}^3=C_5$, $C_{15}^4=C_{15}^2C_{15}^2$, $C_{15}^5=C_{3}$, $C_{15}^6=C_{15}^2C_{15}^2C_{15}^2$, $C_{15}^7=C_{3}C_{15}^2, C_{15}^8=C_{3}C_{5},\cdots$ and hence,
\begin{align}
    {\cal G}^{15} =& \left\{ {\cal G}^{3} {\cal G}^{5} \right\} 
    \nonumber \\
    =& \left\{ C_{15}^5, C_{15}^{10},  C_{15}^{15} \equiv E, \right. \nonumber \\ & \left. \phantom{[} C_{15}^3, C_{15}^6, C_{15}^9, C_{15}^{12}, C_{15}^{15} \equiv E, \right. \nonumber \\ & \left. \phantom{[} C_{15}^5 C_{15}^3 \equiv C_{15}^8, C_{15}^5 C_{15}^6 \equiv C_{15}^{11}, C_{15}^5 C_{15}^9 \equiv C_{15}^{14}, \right. \nonumber \\ & \left. \phantom{[} C_{15}^5 C_{15}^{12} \equiv C_{15}^{2}, C_{15}^{10} C_{15}^3 \equiv C_{15}^{13}, C_{15}^{10} C_{15}^6 \equiv C_{15}, \right. \nonumber \\ & \left. \phantom{[} C_{15}^{10} C_{15}^9 \equiv C_{15}^{4}, C_{15}^{10} C_{15}^{12} \equiv C_{15}^{7} \right\} \nonumber \\ 
    =& \left\{ C_{15}, C_{15}^2, C_{15}^3, C_{15}^4, C_{15}^5, \cdots   E \right\}
    \label{CRT-15-2}
\end{align}
Similarly, it can also be shown that 
\begin{align}
    {\cal G}^{21} =& \left\{{\cal G}^{3} {\cal G}^{7} \right\}     \nonumber \\
    =& \left\{ C_{21}^7, C_{21}^{14},  C_{21}^{21} \equiv E, \right. \nonumber \\ & \left. \phantom{[} C_{21}^3, C_{21}^6, C_{21}^9, C_{21}^{12}, C_{21}^{15}, C_{21}^{18}, C_{21}^{21} \equiv E, \right. \nonumber \\ & \left. \phantom{[} C_{21}^7 C_{21}^3 \equiv C_{21}^{10}, C_{21}^7 C_{21}^6 \equiv C_{21}^{13}, C_{21}^7 C_{21}^9 \equiv C_{21}^{16}, \right. \nonumber \\ & \left. \phantom{[} C_{21}^7 C_{21}^{12} \equiv C_{21}^{19}, C_{21}^7 C_{21}^{15} \equiv C_{21}, C_{21}^7 C_{21}^{18} \equiv C_{21}^{4}, \right. \nonumber \\ & \left. \phantom{[} C_{21}^{14} C_{21}^3 \equiv C_{21}^{17}, C_{21}^{14} C_{21}^6 \equiv C_{21}^{20}, C_{21}^{14} C_{21}^9 \equiv C_{21}^{2}, \right. \nonumber \\ & \left. \phantom{[} C_{21}^{14} C_{21}^{12} \equiv C_{21}^{5}, C_{21}^{14} C_{21}^{15} \equiv C_{21}^8, C_{21}^{14} C_{21}^{18} \equiv C_{21}^{11}\right\} \nonumber \\ 
    =& \left\{ C_{21}, C_{21}^2, C_{21}^3, C_{21}^4, C_{21}^5, \cdots   E \right\}
\end{align}

Insofar as mapping these groups to actual physical systems such as molecules, and condensed phase systems, unlike hexahydroxybenzene, things become more complicated. While it is significantly harder to find molecules that contain such high symmetries, as we discuss later, using Bloch's theorem, we may create condensed phase assemblies and quantum corrals\cite{RevModPhys.quantum-corrals}, the latter being standing wave patterns of electron density created using scanning tunneling microscopy on atomically clean metal surfaces.  
These synthetic systems may contain information relevant to the symmetry within larger numbers. Additionally, future publications will evaluate the use of symmetry within optical beams for this purpose. 
Next, we introduce projection operators that utilize the symmetries within ${\cal G}^{b_i}$ to capture the symmetry of functions within ${\cal G}^{N}$. This turns out to be critical before we embark upon a full discussion of hidden subgroup problems and Shor's algorithm, based on these ideas.

\section{Symmetry-Adapted Functions on Cyclic Groups}
\label{sec:two}

Group theory provides mathematical constructs that allow us to probe the structure of the group and functions that transform according to the symmetry of the group. This paper eventually develops functions that transform according to the groups discussed above, that is ${\cal G}^N \equiv \mathbb{Z}_N$, and then develops physical systems that may contain functions that honor this symmetry. Towards this effort, 
given any group, ${\cal G}$, with members represented as $R$, there exists a set of projection operators defined by\cite{Tinkham},
\begin{equation}
{\cal P}_{\kappa \kappa}^{(j)} = \frac{l_j}{h} \sum_{R \in {\cal G}}
\Gamma_{\kappa\kappa}^{(j)}({R}) P_{R},
\label{proj1}
\end{equation}
where $\left\{ \Gamma^{(j)}({R}) \right\}$ are the irreducible representation matrices of this group. The quantity, $\Gamma_{\kappa\lambda}^{(j)}({R})$ is the $(\kappa,\lambda)$-th
element in the representation matrix of $R$ within the $j$-th irreducible
representation of ${\cal G}$. The $j$-th irreducible
representation has dimensionality $l_j$ and ${\sum_j} {l_j}^2 = h$, with $h$ being the total
number of operations within ${\cal G}$. 

Let us now introduce functions that transform according to the symmetry of the group. Let $f^{[j]}_{\lambda}$ be a function that transforms according to the
$\lambda$-th column of the $j$-th irreducible representation of
${\cal G}$. Using this function, we can define the operation $P_{R}$ in Eq. (\ref{proj1}) using Wigner's notation convention\cite{Tinkham}:
\begin{equation}
f^{[j]}_{\lambda}(R^{-1}x) = P_R f^{[j]}_{\lambda}
(x),
\label{wignerconn}
\end{equation}
Thus the sets $\{ R \}$ and $\{ P_R \}$ are isomorphic where the operations $R$ act on individual spatial elements, where as $P_R$ act on functional elements. 

The ``Great Orthogonality Theorem''\cite{Tinkham} of group theory states that
\begin{align}
    \frac{\sqrt{l_j l_{j^\prime}}}{h} \sum_{R \in {\cal G}} \Gamma_{\kappa\lambda}^{(j)}({R})^* \; \Gamma_{\kappa^\prime\lambda^\prime}^{(j^\prime)}({R}) = \delta_{j,j^\prime} \delta_{\kappa,\kappa^\prime} \delta_{\lambda,\lambda^\prime}
    \label{GOT}
\end{align}
and maintains that the matrices $\Gamma^{(j)}({R})$ form a complete and orthonormal space. In Eq. (\ref{GOT}), $\Gamma_{\kappa\lambda}^{(j)}({R})^*$ indicates the complex conjugate of $\Gamma_{\kappa\lambda}^{(j)}({R})$. 
For Abelian groups the irreducible representations are always one-dimensional and these matrices simply become complex numbers and Eq. (\ref{proj1}) reduces to
\begin{equation}
{\cal P}^{(j)} = \frac{1}{h} \sum_{R \in {\cal G}}
\Gamma^{(j)}({R}) P_{R},
\label{proj1-1D}
\end{equation}
and the orthogonality expression in Eq. (\ref{GOT}) simply becomes
\begin{align}
    \frac{1}{h} \sum_{R \in {\cal G}} \Gamma^{(j)}({R})^* \; \Gamma^{(j^\prime)}({R}) = \delta_{j,j^\prime} 
\end{align}
Consequently, since
\begin{align}
{\cal P}^{(j)} f(x) &= \frac{1}{h} \sum_{R \in {\cal G}}
\Gamma_{}^{(j)}({R}) P_{R} f(x) \nonumber \\ &= \frac{1}{h} \sum_{R \in {\cal G}}
\Gamma_{}^{(j)}({R}) f(R^{-1}x) 
\label{proj1-1D-f}
\end{align}
and since the projection operators resolve the identity, any function can be written as 
\begin{align}
f(x) &= \sum_j {\cal P}^{(j)} f(x) \nonumber \\ &= \frac{1}{h} \sum_{R \in {\cal G}}
\left( \sum_j \Gamma^{(j)}({R}) \right) P_{R} f(x) \nonumber \\ &= \frac{1}{h} \sum_{R \in {\cal G}}
\left( \sum_j \Gamma^{(j)}({R}) \right) f(R^{-1}x) 
\label{proj1-1D-f2}
\end{align}
This can of course also be done for non-Abelian groups where the projections are given by Eq. (\ref{proj1}). 
But cyclic groups are Abelian since all operations within the group commute with each other\cite{Tinkham}. As a result the $\Gamma^{(j)}({R})$ values are scalar and for the group ${\cal G}^N$ in Eq. (\ref{GN-group}), 
\begin{align}
    \Gamma^{(j)}({C_N^k}) = \exp{\imath (2\pi/N)jk}
\end{align}
and
\begin{align}
f_{{\cal G}_N}^{[j]}(x) = {\cal P}^{(j)} f(x) &= \frac{1}{N} \sum_{k=1}^N
\exp{\imath (2\pi/N)jk} f(C_N^{N-k} x) 
\label{proj1-1D-f-GN}
\end{align}
The intuition for this equation is that the circle is divided into $N$ sections (for example, see Figure \ref{fig:G-6-1}), where each section is labeled using the symbol $k$ in the above equation, the prefactor, $\exp{\imath (2\pi/N)jk}$ is periodic with respect to increments in $k$, that is, $\exp{\imath (2\pi/N)j(k+N)} = \exp{\imath (2\pi/N)jk}$ for $j \in [1, \cdots, N]$, and finally, $f(x)$ on the right side above is simply translated identically from one slice to the next and gets multiplied by the respective phase. 
For example, the group ${\cal G}^6 \equiv \left\{ C_6^k \; \vert \; \forall k=1, \cdots , 6 \right\}$ has irreducible representations given by $\Gamma^{[j]}(C_6^k) = \exp{\imath (2\pi/6)jk}$ for $j,k \in [1,\cdots,6]$ and functions that transform according to the symmetry of ${\cal G}^6$ may be written as
\begin{align}
f_{{\cal G}_6}^{[j]}(x) = {\cal P}^{(j)} f(x) &= \frac{1}{6} \sum_{k=1}^6
\exp{\imath (2\pi/6)jk} f(C_6^{6-k} x) 
\label{proj1-1D-f}
\end{align}
where $C_6^{6-k} = \left\{C_6^{k}\right\}^{-1}$ and $j=1\cdots 6$. These functions in fact provide a set of symmatry adapted linear combination of atomic orbtals (SALCs)\cite{Tinkham,cotton} and hence molecular orbitals for hexahydroxybenzene. 

Additionally, as seen in Eq. (\ref{CRT-6}), these can also be written using the representations for ${\cal G}^2$ and ${\cal G}^3$ since $2$ and $3$ are the prime factors of 6 and the group, ${\cal G}^6$, can be written as a direct product of ${\cal G}^2$ and ${\cal G}^3$. Thus, the representations for ${\cal G}^6$ can also be obtained from ${\cal G}^2$ and ${\cal G}^3$ and contain information about its prime factors:
\begin{align}
f_{{\cal G}_2}^{[j]}(x) &= {\cal P}^{(j)} f(x) 
\nonumber \\ &= \frac{1}{2} \left[ 
\exp{\imath j \pi} f(C_2 x) + \exp{\imath 2j \pi} f(x) \right]
\label{proj1-1D-f-GN=2}
\end{align}
\begin{align}
f_{{\cal G}_3}^{[j]}(x) &= {\cal P}^{(j)} f(x) 
\nonumber \\ &= \frac{1}{3} \left[ 
\exp{\imath (2\pi/3)j} f(C_3^2 x) + \right. \nonumber \\ & \left. \phantom{=\frac{1}{3} []} \exp{\imath (2\pi/3)2j} f(C_3 x) + \right. \nonumber \\ & \left. \phantom{=\frac{1}{3} []} \exp{\imath (2\pi/3)3j} f(x) \right]
\label{proj1-1D-f-GN=3}
\end{align}
the product of which yield Eq. (\ref{proj1-1D-f}). Thus in general,
\begin{align}
{\cal P}^{(j)} f(x) &= \bigotimes_i {\cal P}_{b_i}^{(j_i)} f( x)  \nonumber \\ &= \prod_i  \frac{1}{h_{b_i}} \sum_{R \in {\cal G}^{b_i}}
\Gamma_{}^{(j_{b_i})}({R}) f(R^{-1}x)
\label{proj1-1D-f-directproduct}
\end{align}
where we have written the overall projection operator in terms of direct product of projections obtained from the prime factor groups. The prime factor groups have irreducible representations represented using the symbols $\Gamma_{}^{(j_{b_i})}({R})$. Thus, functions that transform according to the symmetry of ${\cal G}$ do contain information regarding the prime factor subgroups of ${\cal G}$.

Since the cyclic groups are Abelian, the discussion in this section has been specific to Abelian groups. However, this discussion can easily be generalized to non-Abelian groups as well. 

\section{Shor's Algorithm as a Hidden Subgroup Construction}
\label{sec:three}
The section above discusses the group structure of factoring, but does not tell us how factoring is to be done based on this information. The section does not show how the prime factors $\{ b_i\}$ can be found, except that functions that transform according to ${\cal G}^{N}$, contain the irreducible representations of $\left\{ {\cal G}^{b_i} \right\}$ within it. 
In this section we will use the methods in the previous section to reconstruct Shor's algorithm. Our goal will be to introduce functions that bear the needed symmetry. 

\subsection{The Extended Cyclic Group ${\cal G}^{N,a}$}
Now consider a number ``$a$", which is a co-prime number of $N$. Based on these two numbers, we introduce a rotation angle $R_z(2\pi/aN) \equiv C_{aN}$. So $C_{aN}$ is a representation for the noted rotation angle about the $z$-axis on a two-dimensional circle on the $xy$-plane. In addition, we define the rotation operations $C_a \equiv C_{aN}^N = R_z(2\pi/a)$, and $C_N \equiv  C_{aN}^a = R_z(2\pi/N)$ and associated with these we introduce a group with elements given by
\begin{align}
    {\cal G}^{N,a} = \left\{ C_{aN}, C_{aN}^2, \cdots, C_{aN}^{a}, \cdots, C_{aN}^N, \cdots, C_{aN}^{aN-1}, E \right\}
    \label{Group-GaN}
\end{align}
where $E=C_{aN}^{aN}$, is the identity operation, and we have individually identified the operations $C_N$ and $C_a$ within ${\cal G}^{N,a}$. It is trivial to show that these elements form a group because, the group is closed under rotations, all elements contain an inverse and the identity operation is also present in the group. However, we have now developed a {\em finer}-grade representation of ${\cal G}^{N}$. Thus, in essence, each element of the original group ${\cal G}^{N}$ is now further divided into $a$ segments. 

\begin{figure}
    \centering
    \subfigure[]{\includegraphics[width=0.32\linewidth]{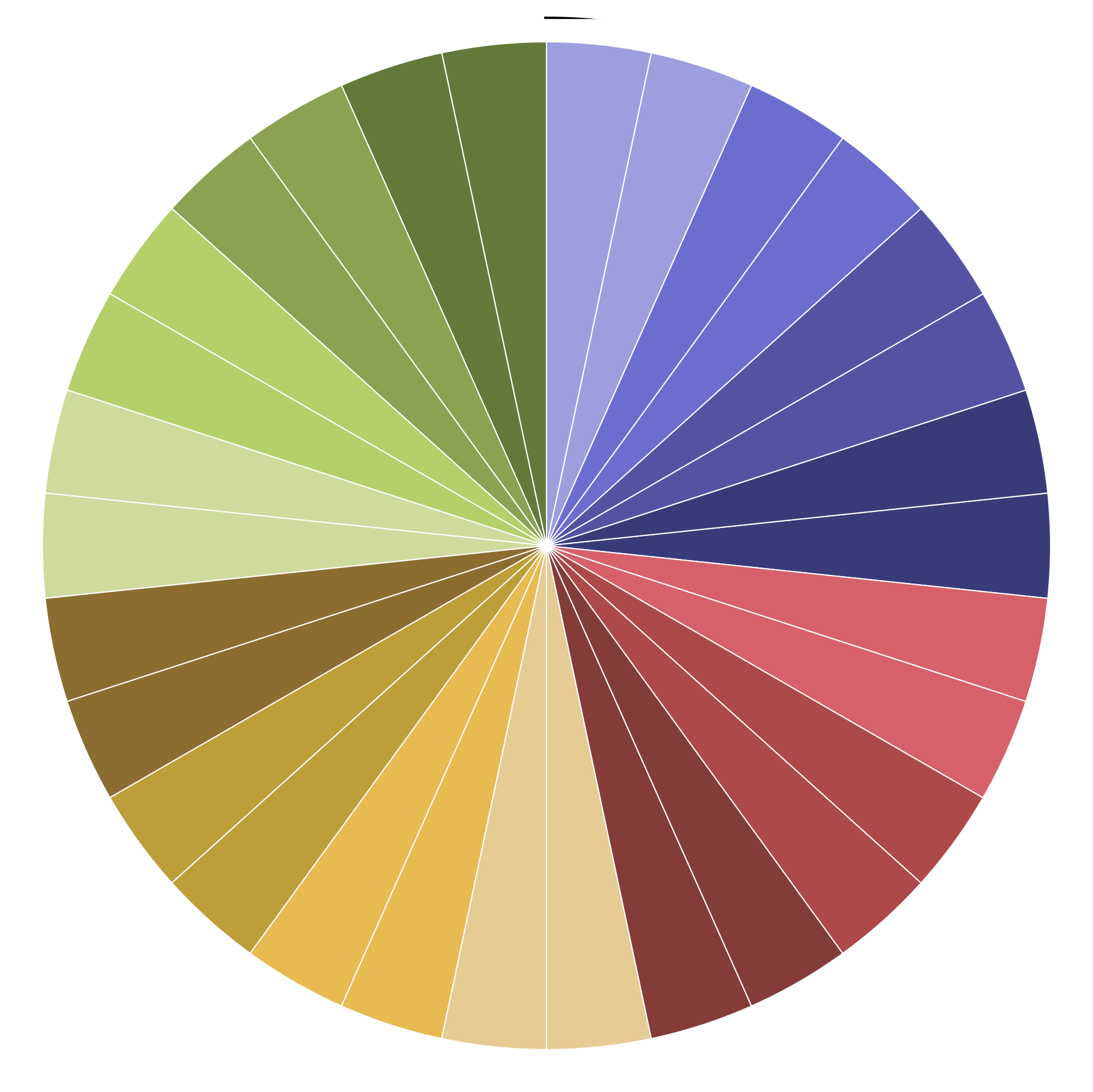}}
    \subfigure[]{\includegraphics[width=0.32\linewidth]{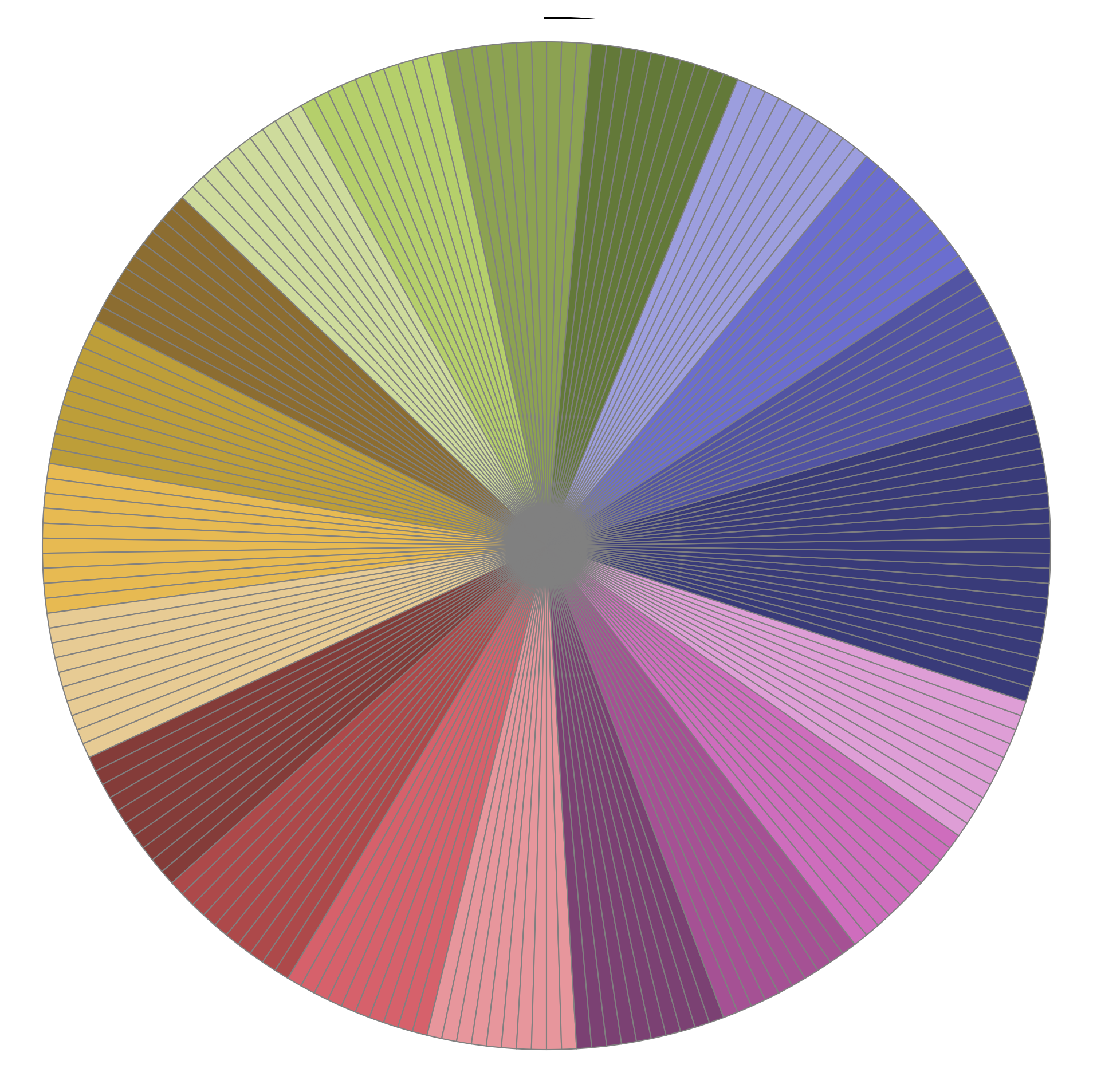}}
    \subfigure[]{\includegraphics[width=0.32\linewidth]{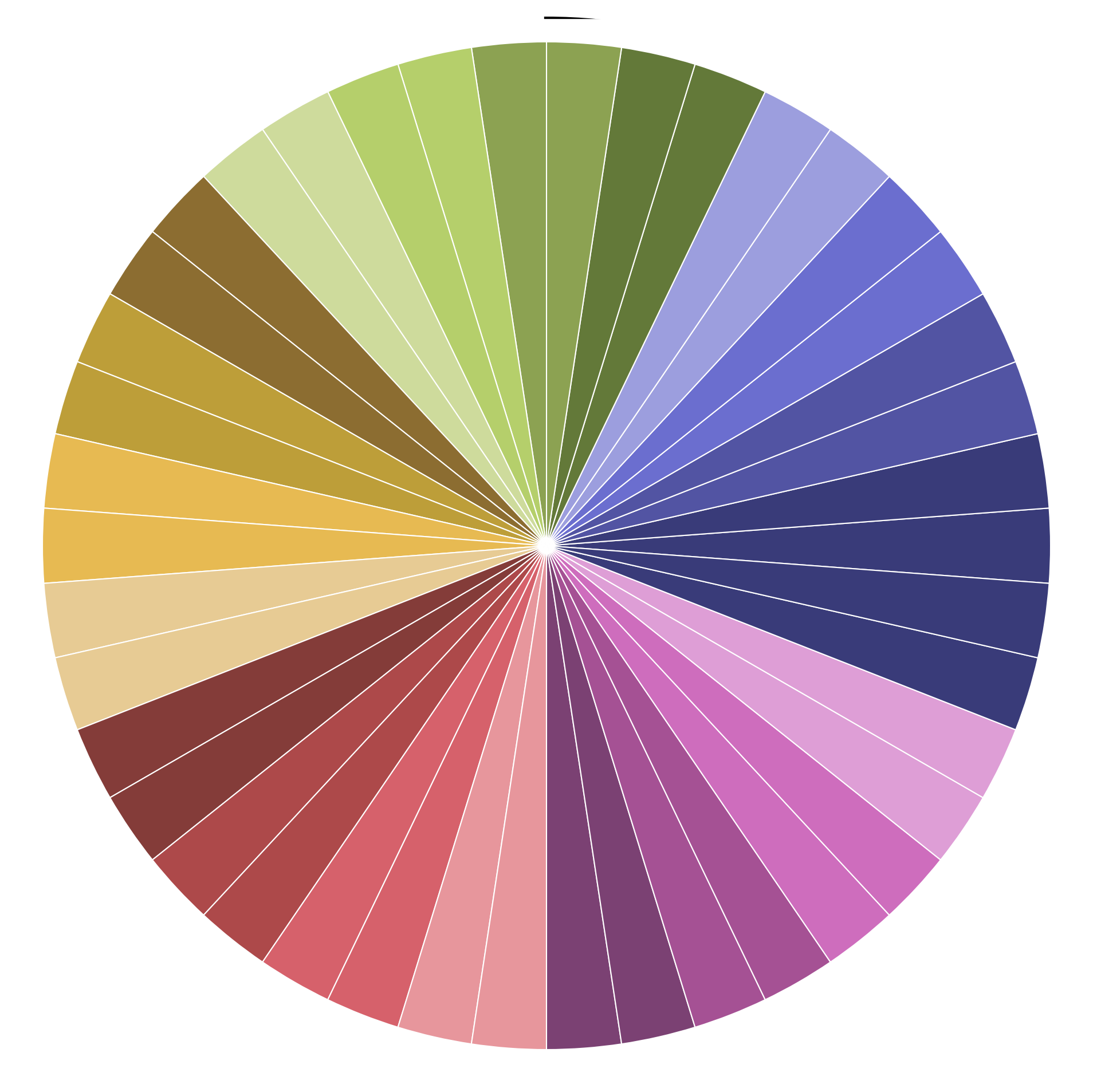}}
    \caption{Illustration  of (a): ${\cal G}^{15,2}$, (b): ${\cal G}^{21,10}$ and (c): ${\cal G}^{21,2}$. In Figure (c) we have chose $a=2$, unlike in Eq. (\ref{G-21-10}). In each figure, the circle is first divided into $N$ equal portions and then each slice is further subdivided in $a$ parts. The colors are chosen purely to make the regions distinct. The isomorphic objects being constructed here are symmetric with rotations, and in Sections \ref{HSG_Groups} and \ref{HSG_Groups-2} we discuss functions defined on these objects. Also see Figure \ref{fig:G-6}. (These images were generated using the AI tool Microsoft Copilot.)}
    \label{fig:G-15-21}
\end{figure}
\begin{figure}
    \centering
    \subfigure[]{\includegraphics[width=0.32\linewidth]{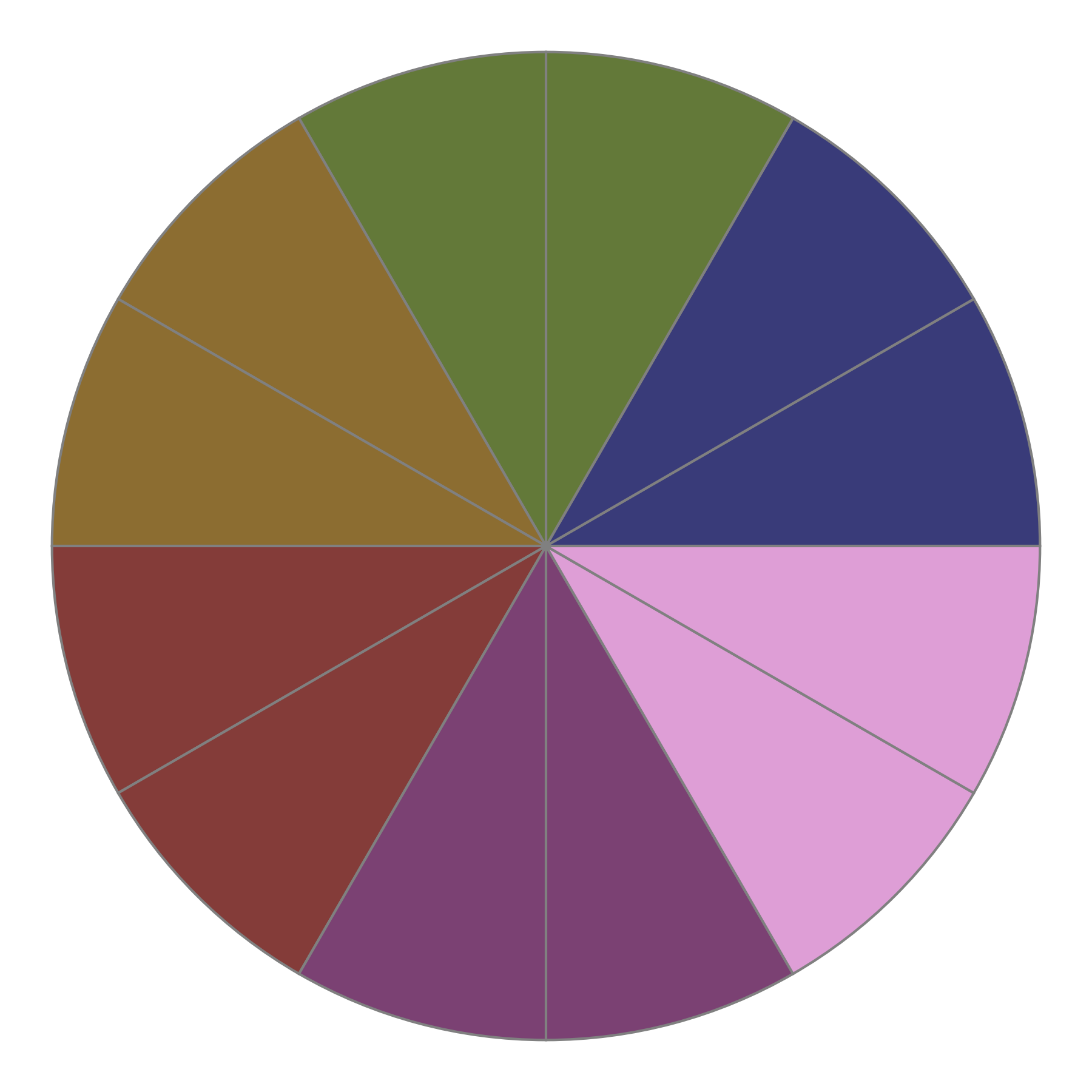}}
    \subfigure[]{\includegraphics[width=0.32\linewidth]{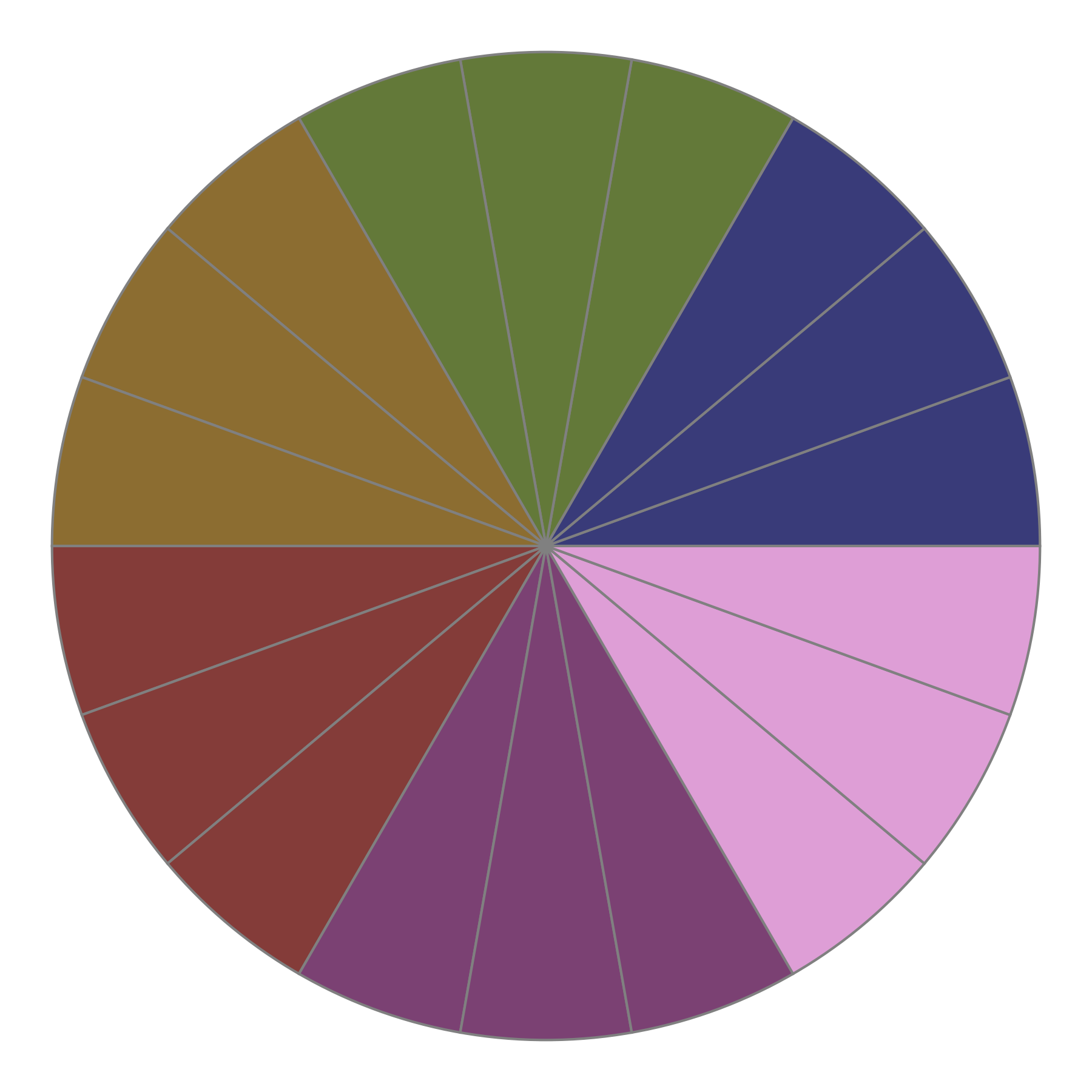}}
    \subfigure[]{\includegraphics[width=0.32\linewidth]{Figures/Benzenehexol.png}}    
    \caption{Similar to Figure \ref{fig:G-15-21}, but figures (a) and (b) are ${\cal G}^{6,2}$ and ${\cal G}^{6,3}$, respectively. Figure (c) is the molecule hexabydroxybenzene (or benzenehexol), with the OH groups pointing out of the plane, that has the same symmetry as ${\cal G}^{6}$, which, as shown in Sections \ref{HSG_Groups} is a hidden subgroup of ${\cal G}^{6,2}$ and ${\cal G}^{6,3}$. (Figures (a) and (b) were generated using the AI tool Microsoft Copilot.)}
    \label{fig:G-6}
\end{figure}
To illustrate this group, we have presented Figures \ref{fig:G-15-21} and \ref{fig:G-6} that are further discussed below. Let us consider $N=15$ and $a=2$. The rotation increment is $R_z(2\pi/30) \equiv C_{30}$ and the corresponding rotation operations that belong to the group are $C_{2}=C_{30}^{15}=R_z(\pi)$, and corresponds to the integer $15 \text{ (that is}, N)$, and $C_{15}=C_{30}^2=R_z(2\pi/15)$, and corresponds to the integer $2 \text{ (that is, } a)$. The elements of such a group then are
\begin{align}
    {\cal G}^{15,2} = \left\{ C_{30}, C_{30}^2, \cdots, C_{30}^{15}, \cdots, E \right\}
    \label{G-15-2}
\end{align}
and each element corrsponds to an integer in the range $[1,30]$. The group ${\cal G}^{15,2}$ is illustrated through Figure \ref{fig:G-15-21}(a), where there are 15 main slices to the circle and then each slice is further divided into two parts. For N=21 and $a=10$,
\begin{align}
    {\cal G}^{21,10} = \left\{ C_{210}, C_{210}^2, \cdots, C_{210}^{10}, \cdots, C_{210}^{21}, \cdots, E \right\}
    \label{G-21-10}
\end{align}
and is illustrated in Figure \ref{fig:G-15-21}(b). Also see Figure \ref{fig:G-6}. All of these groups are closed under rotation, individual elements have inverse operations and there is an identity in the group as well. We will show below how Shor's algorithm corresponds to symmetries inside this group. 

\subsection{Hidden Subgroup Formulation of Shor: subgroup ${\cal G}^{N} \in {\cal G}^{N,a}$}
\label{HSG_Groups}
We can rewrite Eq. (\ref{Group-GaN}) using cosets of ${\cal G}^{N}$ within ${\cal G}^{N,a}$, 
\begin{align}
    {\cal G}^{N,a} &= \left\{ C_{aN}^{i} \left\{ C_{N}, C_{N}^2, \cdots, C_{N}^{N-1}, E \right\} \vert \; \forall \; i \in [0,a-1] \right\} \nonumber \\ &= \left\{ C_{aN}^{i}  {\cal G}^{N} \vert \; \forall \; i \in [0,a-1] \right\}
    \label{Group-GaN-sub}
\end{align}
Thus, the rotation operations in ${\cal G}^{N}$ are each divided into a further set of $a$ increments each to arrive at ${\cal G}^{N,a}$. 
We also note that
\begin{align}
\forall \; C_{aN}^{i} &\in {\cal G}^{N,a} ; i=[0,aN-1] \\
C_{aN}^{i} {\cal G}^{N} &= C_{aN}^{i+ja} {\cal G}^{N}; \; \forall j=[0,N-1]
\label{coset-GN}
\end{align}
since $C_{aN}^{i+ja} = C_{aN}^{i} {C_{N}^{j}}$.
Hence the cosets of ${\cal G}^{N}$ cyclically repeat within ${\cal G}^{N,a}$. This kind of a substructure is only possible because $N$ and $a$ are coprime numbers (or it can also be constructed by considering the largest common divisor of $N$ and $a$ as the fundamental angle increment, but that then loses the power of encoding every single integer onto the rotation group). The fact that these are coprimes allows all natural numbers to be represented inside the group ${\cal G}^{N,a}$. As we see from Eqs. (\ref{GN-group}) and (\ref{Group-GaN-sub}) the prime factors of $N$ do also appear within ${\cal G}^{N,a}$ which as we will see below is a key requirement made possible because $N$ and $a$ are coprime numbers.

\begin{figure}
    \centering
    \subfigure[]{\includegraphics[width=0.32\linewidth]{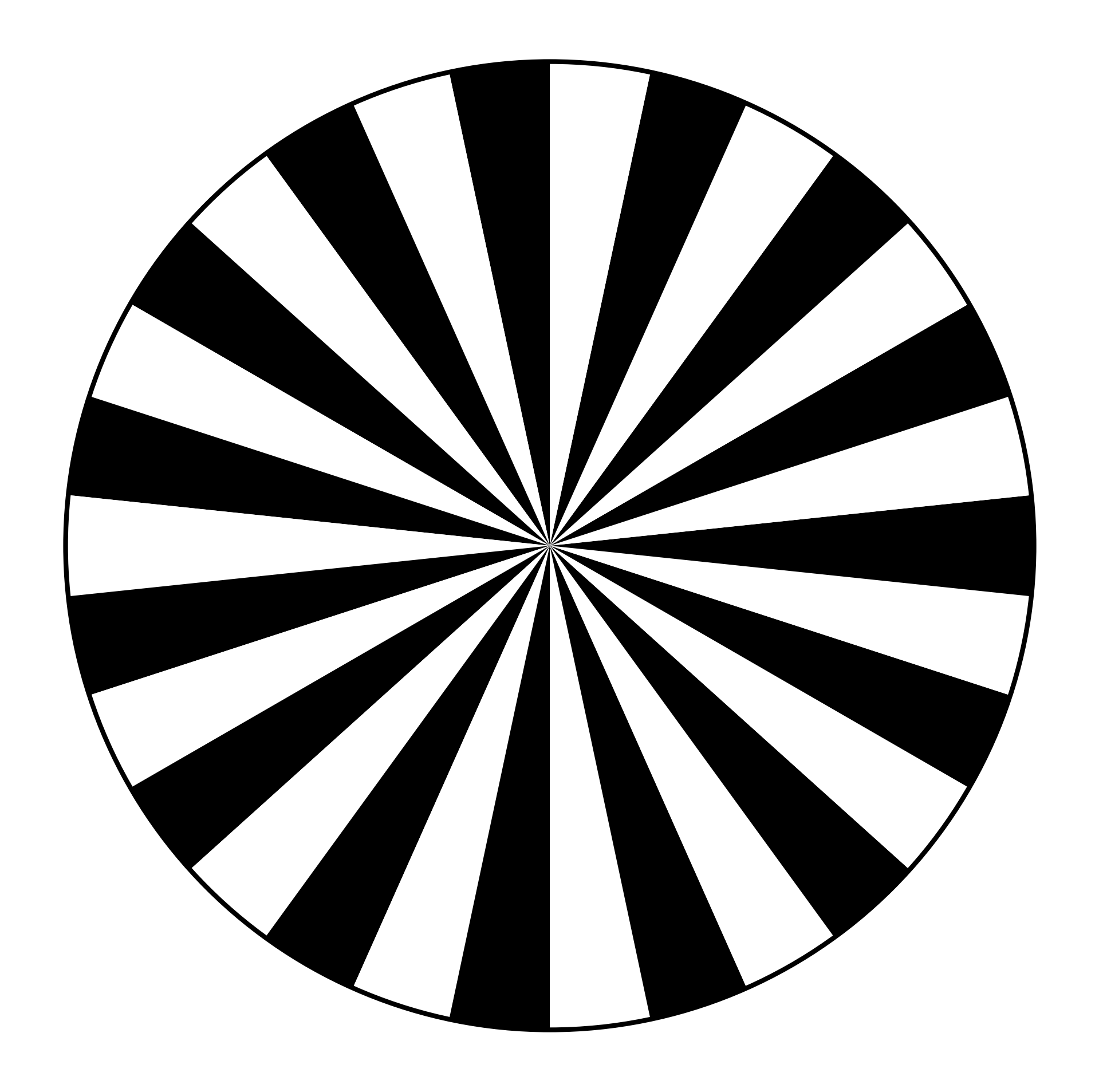}}
    \subfigure[]{\includegraphics[width=0.32\linewidth]{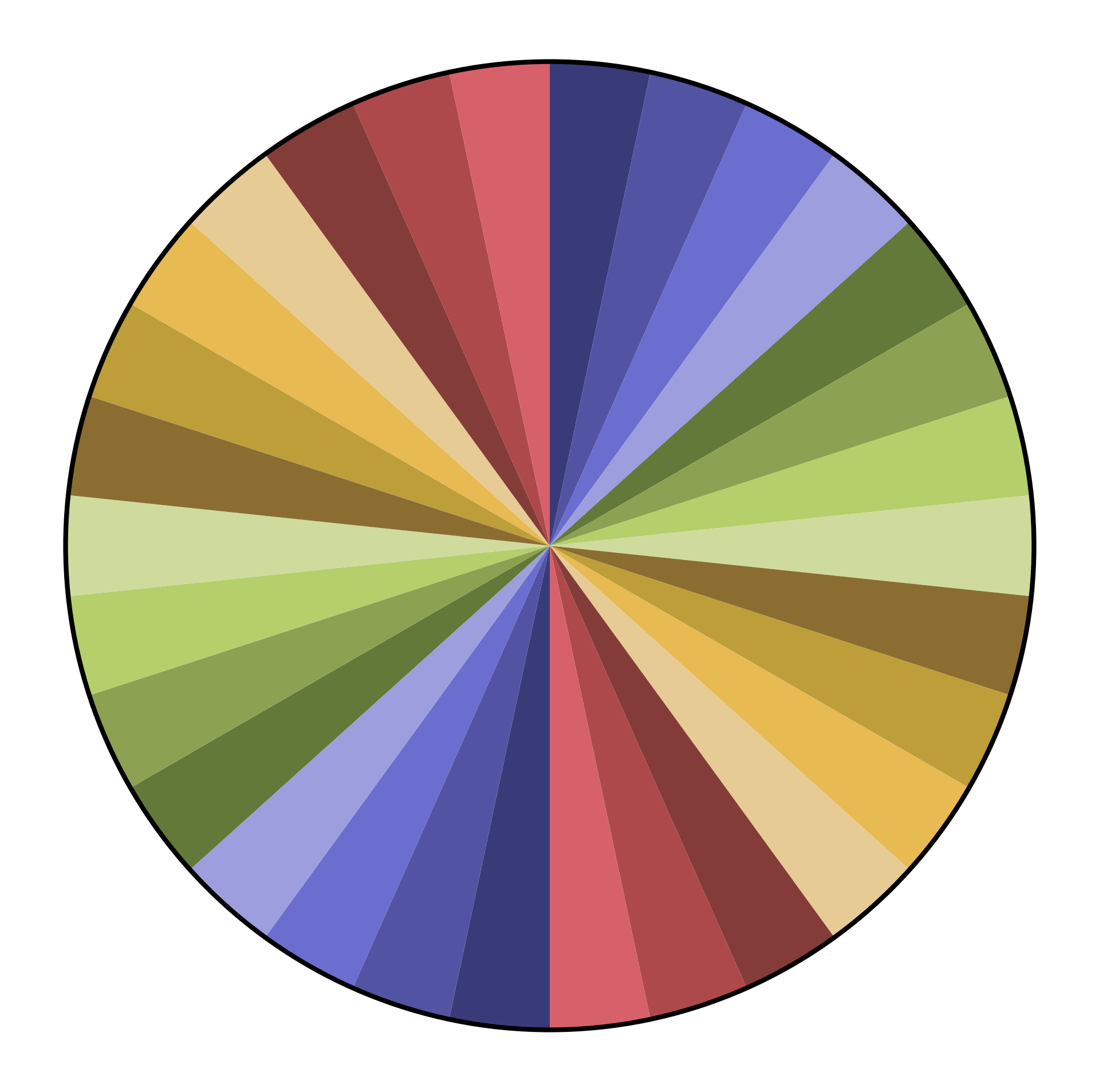}}
    \caption{Illustration  of (a): Eq. (\ref{Group-Ga-GaN-sub}) and (b): Eq. (\ref{Group-GaN-a-sub}) for the group ${\cal G}^{15,2}$. In both cases, the cosets generated by $C_{aN}^{i}  {\cal G}^{N}$ (Figure (a)) and $C_{aN}^{i}  {\cal G}^{a}$ (Figure (b)) are depicted using the same color. For example, in Figure (a), the action of all $C_{aN}^{i+ja}; \; \forall j=[0,N-1]$, on $ {\cal G}^{N}$, for odd or even values of $i$ yields either the black or the white regions. Likewise in Figure (b) the action of all $C_{aN}^{i+jN}; \; \forall j=[0,a-1]$, on $ {\cal G}^{a}$, yields each region with a specific color. As we show here, functions such as $f(x) = a^x \pmod{N}$ (also see Section \ref{sec:five}) have the same values inside the, thus created cosets, and hence hide ${\cal G}^{N}$. (These images were generated using the AI tool Microsoft Copilot.)}
    \label{fig:Coset-G-15}
\end{figure}
Additionally, we may also write
Eq. (\ref{Group-GaN}) using cosets of ${\cal G}^{a}$ within ${\cal G}^{N,a}$. That is, 
\begin{align}
    {\cal G}^{N,a} &= \left\{ C_{aN}^{i} \left\{ C_{a}, C_{a}^2, \cdots, C_{a}^{a-1}, E \right\} \vert \; \forall \; i \in [0,N-1] \right\} \nonumber \\ &= \left\{ C_{aN}^{i}  {\cal G}^{a} \vert \; \forall \; i \in [0,N-1] \right\}
    \label{Group-Ga-GaN-sub}
\end{align}
and
\begin{align}
\forall \; C_{aN}^{i} &\in {\cal G}^{N,a} ; i=[0,aN-1] \\
C_{aN}^{i} {\cal G}^{a} &= C_{aN}^{i+jN} {\cal G}^{a}; \; \forall j=[0,a-1] 
\label{coset-Ga}
\end{align}
Hence, as before the cosets of ${\cal G}^{a}$ also cyclically repeat within ${\cal G}^{N,a}$. 
Equations (\ref{coset-GN}) and (\ref{coset-Ga}) are also described pictorially in Figure \ref{fig:Coset-G-15} for $N=15$ and $a=2$. In Figure \ref{fig:Coset-G-15}(a), the set of all black (or white) slices independently represent elements of ${\cal G}^{N}$. That is, there are two copies of ${\cal G}^{N}$ in ${\cal G}^{N,a}$ due to the fact that $a$ was chosen to be 2. Upon action of $C_{aN}^{i}$ on the black slice set, we recover the white slice set. This action is represented by Eq. (\ref{coset-GN}). Similarly, in Figure \ref{fig:Coset-G-15}(b), the 30 slices corresponding to ${\cal G}^{N,a} \equiv {\cal G}^{15,2}$ are divided into 15 different sets of  ${\cal G}^{a}$ where each set contains diametrically opposite slices that are marked with the same color. That is there are 15 copies of ${\cal G}^{a}$ in ${\cal G}^{N,a}$ for $N=15$ and $a=2$. Here again, upon action of $C_{aN}^{i}$ on any of these sets with the same color, yields a different set of the same color and this action is represented by Eq. (\ref{coset-Ga}).

Furthermore, by choosing $a$ as a prime factor of $N$, we have assured that there are no additional prime numbers that factor ${\cal G}^{N,a}$ beyond the $\left\{ b_i \right\}$ discussed earlier, in the sense of the Chinese Remainder Theorem. That is, the prime factor sub-groups ${\cal G}^{b_i}$ of ${\cal G}^N$ also remain the prime factor sub-groups ${\cal G}^{N,a}$ which can also be created by powers of these subgroups. As we will see below, the specific construction of ${\cal G}^{N,a}$ in this section makes it possible to develop ${\cal G}^{N}$ (and ${\cal G}^{a}$) as hidden subgroups of ${\cal G}^{N,a}$. 

\subsection{Functions that transform according to the symmetry of ${\cal G}^{N,a}$}
A function such as $f(x) = a^x \pmod{N}$, which is used in Shor's algorithm, may be understood as follows within the group representation that we have constructed here. We first note that the values of $x$ are mapped to the integer values,
\begin{align}
x\rightarrow (i,j)
\end{align}
where, $x=i+jN$ and as per Eq. (\ref{Group-Ga-GaN-sub}), $x$ may be mapped to a single slice in ${\cal G}^{N,a}$ since 
\begin{align}
{\cal G}^{N,a} &= \left\{ C_{aN}^{i} C_{a}^j \vert \; \forall \; i \in [0,N-1]; j \in [0,a-1] \right\}
\label{Group-GaN-a-sub}
\end{align}
As per Figure \ref{fig:Coset-G-15}(b), the integer $j$ represents a specific color (pair of slices) whereas $i$ represents the incremental angle to rotate between colors. However, we may also map $x$ to a different set of indices if using the cosets of ${\cal G}^{N}$, that is
\begin{align}
x\rightarrow (i^\prime,j^\prime)
\end{align}
where, $x=i^\prime+j^\prime a$ and as per Eq. (\ref{Group-GaN-sub}),
\begin{align}
{\cal G}^{N,a} &= \left\{ C_{aN}^{i^\prime} C_{N}^{j^\prime} \vert \; \forall \; i^\prime \in [0,a-1]; j^\prime \in [0,N-1] \right\}
\end{align}
Again, as per Figure \ref{fig:Coset-G-15}(a), the integer $j^\prime$ represents a specific color (black or white) whereas $i^\prime$ represents the incremental angle to rotate between colors.

The outcome of $f(x) = a^x \pmod{N}$ is the value $\beta$ such that $ a^x  = \alpha N + \beta$. That is 
\begin{align}
a^x \rightarrow (\alpha,\beta)
\end{align}
and, as before, $f(x)$ gets mapped to a single slice in ${\cal G}^{N,a}$ since 
\begin{align}
{\cal G}^{N,a} &= \left\{ C_{aN}^{\beta} C_{a}^\alpha \vert \; \forall \; \beta \in [0,N-1]; \alpha \in [0,a-1] \right\}
\label{axmodn-coset}
\end{align}
or the quantity $C_{aN}^{\beta} {\cal G}^{a}$ specifies all elements of a coset (for different values, $\alpha$) of the subgroup ${\cal G}^{a}$ within ${\cal G}^{N,a}$. Thus elements of the coset $C_{aN}^{\beta} {\cal G}^{a}$ remain {\em hidden} to $f(x) = a^x \pmod{N}$ and all elements of the coset have exactly the same function value. 

Thus the functional map
\begin{align}
x \rightarrow a^x \pmod{N}
\end{align}
is equivalent to the map between group elements of ${\cal G}^{N,a}$ given by 
\begin{align}
C_{aN}^{i} C_{a}^j \rightarrow C_{aN}^{\beta} C_{a}^\alpha
\end{align}
with 
\begin{align}
i \in [0,N-1]; j \in [0,a-1]; \nonumber \\ \beta \in [0,N-1]; \alpha \in [0,a-1] 
\end{align}
or
\begin{align}
C_{aN}^{i^\prime} C_{N}^{j^\prime} \rightarrow C_{aN}^{\beta} C_{a}^\alpha
\end{align}
with 
\begin{align}
i^\prime \in [0,a-1]; j^\prime \in [0,N-1]; \nonumber \\ \beta \in [0,N-1]; \alpha \in [0,a-1] 
\end{align}
and the result of the modular operation is the number $\beta$. 

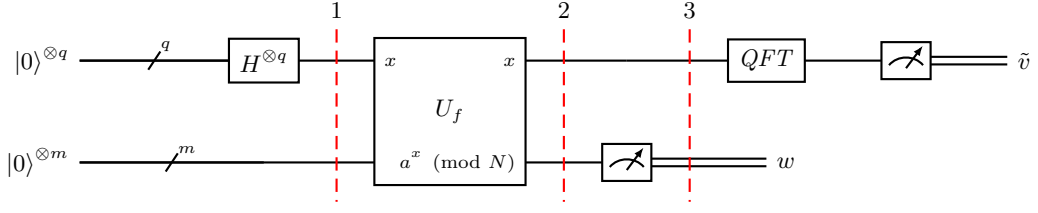
\begin{figure*}[t]
\input{Figures/Shor-ckt}
\caption{\label{fig:shor} Quantum circuit for Shor's algorithm.}
\end{figure*}
Here are a few examples of how this works. Consider $N=15$, $a=2$,
\begin{align}
a^x: x\equiv(i^\prime,j^\prime) &\rightarrow (\alpha,\beta) \nonumber \\
a^0: x\equiv(0,0) &\rightarrow (0,1) \nonumber \\
a^1: x\equiv(1,0) &\rightarrow (0,2) \nonumber \\
a^2: x\equiv(0,1) &\rightarrow (0,4) \nonumber \\
a^3: x\equiv(1,1) &\rightarrow (0,8) \nonumber \\
a^4: x\equiv(0,2) &\rightarrow (1,1) \nonumber \\
a^5: x\equiv(1,2) &\rightarrow (2,2) \nonumber \\
a^6: x\equiv(0,3) &\rightarrow (4,4) \nonumber \\
a^7: x\equiv(1,3) &\rightarrow (8,8) \nonumber \\
a^8: x\equiv(0,4) &\rightarrow (7,1) \nonumber \\
\vdots 
\end{align}
The sequence $1, 2, 4, 8$ repeats with period 4. This period is determined using the quantum 
circuit shown in Figure~\ref{fig:shor}. The top register is initialized in a uniform superposition 
over all $x$, and the oracle $U_f$ then computes $a^x \pmod{N}$ into the bottom register. 
Measuring the bottom register collapses the top register onto a superposition over all $x$ values 
sharing the same residue $\beta$, which is a periodic superposition with the same period $r$ as 
the function. Applying the QFT to the top register converts this periodicity into peaks at integer 
multiples of $aN/r$, from which $r$ is recovered by classical post-processing. The prime factors 
of $N$ then follow from $\gcd(a^{r/2} \pm 1, N)$. In this sense, the core of Shor's algorithm 
is a period-finding routine executed through the quantum Fourier transform.

For $N=21$, $a=10$, 
\begin{align}
a^x: x\equiv(i^\prime,j^\prime) &\rightarrow (\alpha,\beta) \nonumber \\
a^0: x\equiv(0,0) &\rightarrow (0,1) \nonumber \\
a^1: x\equiv(1,0) &\rightarrow (0,10) \nonumber \\
a^2: x\equiv(2,0) &\rightarrow (4,16) \nonumber \\
a^3: x\equiv(3,0) &\rightarrow (47,13) \nonumber \\
a^4: x\equiv(4,0) &\rightarrow (476,4) \nonumber \\
a^5: x\equiv(5,0) &\rightarrow (4761,19) \nonumber \\
a^6: x\equiv(6,0) &\rightarrow (47619,1) \nonumber \\
a^7: x\equiv(7,0) &\rightarrow (476190,10) \nonumber \\
a^8: x\equiv(8,0) &\rightarrow (4761904,16) \nonumber \\
a^9: x\equiv(9,0) &\rightarrow (47619047,13) \nonumber \\
a^{10}: x\equiv(0,1) &\rightarrow (476190476,4) \nonumber \\
a^{11}: x\equiv(1,1) &\rightarrow (4761904761,19) \nonumber \\
a^{12}: x\equiv(2,1) &\rightarrow (47619047619,1) \nonumber \\
a^{13}: x\equiv(3,1) &\rightarrow (476190476190,10) \nonumber \\
\vdots 
\end{align}
and the $\beta$-values: $1, 10, 16, 13, 4, 19$ repeat.

Now, since $\beta$ identifies a specific coset inside ${\cal G}^{N,a}$ as per Eqs. (\ref{Group-Ga-GaN-sub}) and (\ref{axmodn-coset}), the function yields a constant value of $\beta$ for any two values of $x$, say $x_1$ and $x_2$, such that  
\begin{align}
x_1 \rightarrow (i_1,j_1) \equiv C_{aN}^{i_1}  C_{a}^{j_1} \xrightarrow[]{a^{x_1}} C_{aN}^{\beta}  C_{a}^{\alpha_1}
\label{HSB1}
\end{align}
and 
\begin{align}
x_2 \rightarrow (i_2,j_2) \equiv C_{aN}^{i_2}  C_{a}^{j_2} \xrightarrow[]{a^{x_2}} C_{aN}^{\beta}  C_{a}^{\alpha_2}.
\label{HSB2}
\end{align}
or
\begin{align}
x_1 \rightarrow (i^\prime_1,j^\prime_1) \equiv C_{aN}^{i_1^\prime}  C_{N}^{j_1^\prime} \xrightarrow[]{a^{x_1}} C_{aN}^{\beta}  C_{a}^{\alpha_1}
\label{HSB3}
\end{align}
and 
\begin{align}
x_2 \rightarrow (i_2^\prime,j_2^\prime) \equiv C_{aN}^{i_2^\prime}  C_{N}^{j_2^\prime} \xrightarrow[]{a^{x_2}} C_{aN}^{\beta}  C_{a}^{\alpha_2}.
\label{HSB4}
\end{align}
This leads to the repeating cycle above. 
Thus, the function $a^x \pmod{N}$ has a constant value inside a specific coset of ${\cal G}^{a}$. For example, if each coset is generated for a specific value of $\beta$, given by $\left\{ C_{aN}^{\beta}  {\cal G}^{a} \vert \; \beta \in [0,a-1] \right\}$, the different values of $\beta$ correspond to different cosets, and $a^x \pmod{N}$ has a constant value inside a specific coset of ${\cal G}^{a}$. 
Hence ${\cal G}^{a}$ (and ${\cal G}^{N}$) is a {\em hidden subgroup} of ${\cal G}^{N,a}$ as per $f(x) = a^x \pmod{N}$.

\section{Periodic Wavefunctions and Cyclic Symmetry}
\label{sec:four}
\label{HSG_Groups-2}
As noted, Shor's algorithm exploits the hidden subgroup property of ${\cal G}^{N}, {\cal G}^{a} \in {\cal G}^{N,a}$ to find the prime factors of some large integer. The precise statement with respect to Shor's algorithm is that the function $f(x)=a^x \pmod{N}$ is such that, every element of a coset of ${\cal G}^{a}$ has the same value of the function, but elements in different cosets have different function values, 
which is the essence of the hidden subgroup problem\cite{Burton2011} in mathematics. From Eqs. (\ref{HSB1})-(\ref{HSB4}) it is clear that both $x_1$ and $x_2$ lead to the same function value of $\beta$. 

We may now move further and ask if molecular, and condensed phase systems exist or if synthetic systems can be constructed with these same symmetry properties. Towards this, let us consider general functions $f(x)$ with $x \in [0,2\pi]$. We also note that $x$ arises from incremental rotation operations in ${\cal G}^{N}$, that are in turn mapped to integers in $[0,N]$. All this is allowed by the map of integers to the unit circle in the sections above. 
To connect to molecular systems, we may invoke Bloch's theorem\cite{hamermesh} where the solutions to the Schr\"{o}dinger equation in a periodic potential can be expressed as plane waves modulated by periodic functions with higher frequency than the chosen plane-wave, that is,,
\begin{equation}
    \psi_k(x) = \exp{\imath k x} u_k(x)
    \label{bloch-defn}
\end{equation}
where $\exp{\imath k x}$ is periodic with period $2\pi$ (or $N\frac{2\pi}{N}$), $u_k(x)$ is periodic about $\frac{2\pi}{N}$, and $k$ is a whole number here that represents the lattice momentum. In chemistry and condensed matter physics, the wavefunction above is the solution to a Schr\"{o}dinger equation in a periodic potential, with period $\frac{2\pi}{N}$. These ideas are related to point groups in molecules\cite{cotton} and to periodic lattices in condensed matter\cite{Tinkham}.

Comparing Eqs. (\ref{proj1-1D-f-GN}) and (\ref{bloch-defn}), we note that the function $u_k(x)$ in Eq. (\ref{bloch-defn}) represents the concatenated periodically translated piece-wise set $\left\{ f(C_N^{N-k} x) \right\}$ and hence $u_k(x)$ is periodic with respect to $2\pi/N$ rotations. Likewise, $\exp{\imath (2\pi/N)jk}$ in Eq. (\ref{proj1-1D-f-GN}) is periodic about $2\pi$ which is also the case of $\exp{\imath k x}$ in Eq. (\ref{bloch-defn}). Thus Eqs. (\ref{proj1-1D-f-GN}) and (\ref{bloch-defn}) are essentially identical, which implies the state in Eq. (\ref{bloch-defn}) if appropriately created may be used to tackle problems described in the previous section. 

For example, from Eq. (\ref{bloch-defn}), 
\begin{align}
    \psi_k(x+2\pi/N) &= \exp{\imath k (x+2\pi/N)} u_k(x+2\pi/N) \nonumber \\
    &= \exp{\imath k (2\pi/N)} \exp{\imath k x} u_k(x)\nonumber \\
    &= \exp{\imath k (2\pi/N)} \psi_k(x)
    \label{bloch-sym}
\end{align}
and the wavefunction gains a phase $\exp{\imath k (2\pi/N)}$ from $x \rightarrow x+2\pi/N$. Furthermore, $k$ takes integer values $[0,N-1]$ since
\begin{align}
    \psi_k(x+N*(2\pi/N)) &= \exp{\imath k N* (2\pi/N)} \psi_k(x) 
    \label{bloch-sym-2}
\end{align}
and the right side is only equal to $\psi_k(x)$ when $k$ is an integer. The only unique values are for $k \in [0,N-1]$. 
Likewise, from Eq. (\ref{proj1-1D-f-GN})
\begin{align}
f_{{\cal G}_N}^{[j]}(x+&2\pi/N) = f_{{\cal G}_N}^{[j]}(C_N x) = P_{C_N^{-1}} f_{{\cal G}_N}^{[j]}(x) \nonumber \\ 
=& P_{C_N^{-1}} \frac{1}{N} \sum_{k=1}^N
\exp{\imath (2\pi/N)jk} f(C_N^{N-k} x) \nonumber \\ 
=& \frac{1}{N} \sum_{k=1}^N
\exp{\imath (2\pi/N)jk} P_{C_N^{-1}} P_{C_N}^k f(x) \nonumber \\ 
=& \frac{1}{N} \sum_{k=1}^N
\exp{\imath (2\pi/N)jk} P_{C_N}^{k-1} f(x) \nonumber \\ 
=& \frac{1}{N} \sum_{k=1}^N
\exp{\imath (2\pi/N)j(k-1)} \nonumber \\ 
& \phantom{\frac{1}{N} \sum_{k=1}^N} \exp{\imath (2j\pi/N)} f(C_N^{N-(k-1)} x) \nonumber \\ 
=& \exp{\imath (2j\pi/N)} f_{{\cal G}_N}^{[j]}(x)
\label{proj1-1D-f-GN-translate}
\end{align}
Between Eqs. (\ref{bloch-sym}) and (\ref{proj1-1D-f-GN-translate}), we can see that Bloch functions have identical symmetry properties as the irreducible representations for the prime factor groups discussed above. {\em Hence this must also imply the presence of a Chinese Remainder Theorem in molecular and condensed phase systems implying the presence of information pertaining to the prime factors within the states of such systems.} So the  question is, if matter (molecules, lattices, optical devices) can be created with such tailored properties.

\section{Molecular and Condensed-Phase Realizations}
\label{sec:five} 
\subsection{Symmetry-Adapted Linear Combinations of atomic orbitals from prime-factor subgroup representations}
\begin{figure*}
    \centering
    \includegraphics[width=\linewidth]{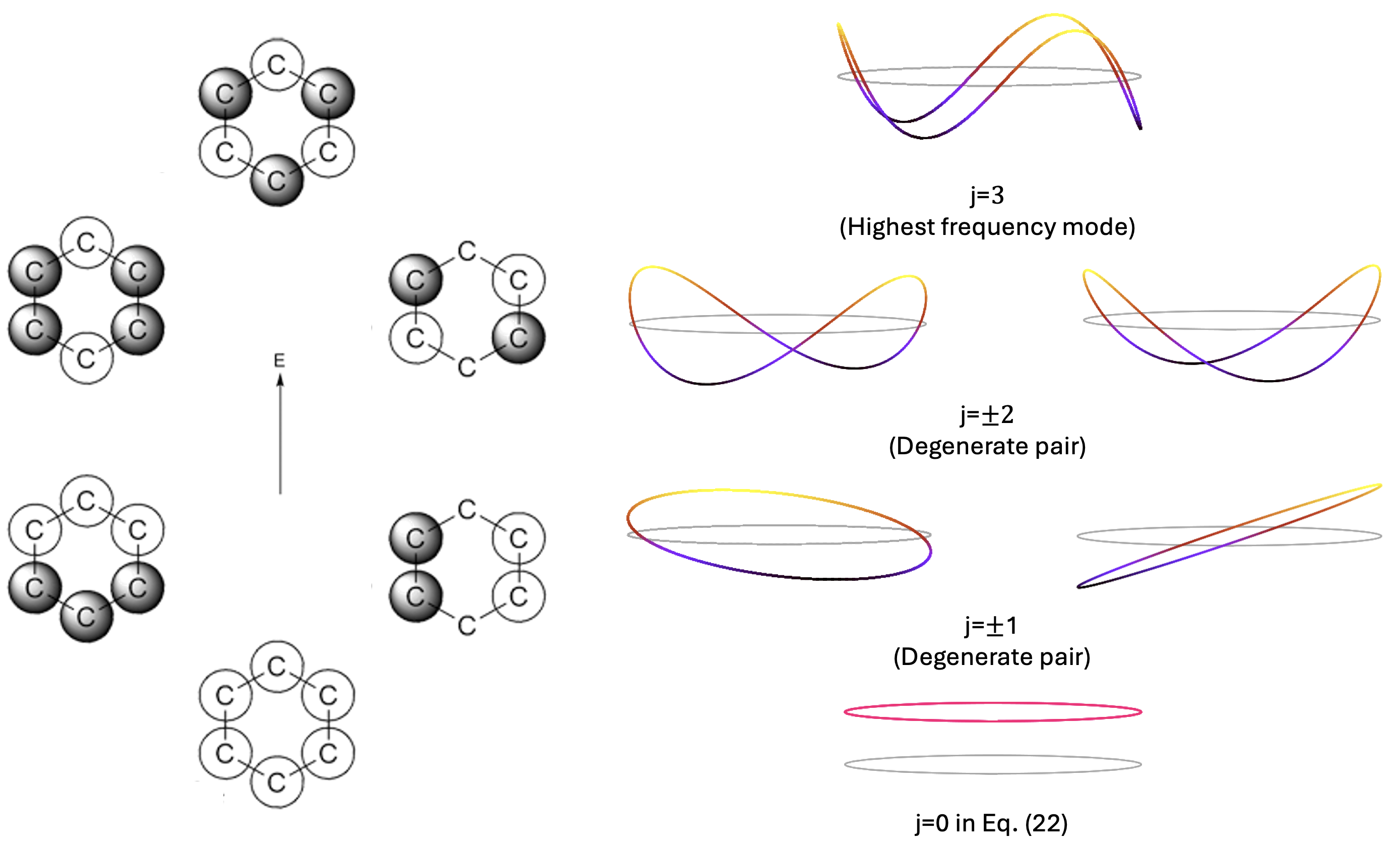}
    \caption{Discrete and continuous representations for Eq. (\ref{proj1-1D-f}). Left: $\pi$-molecular orbitals of benzene (six-membered ring), obtained from linear combinations of atomic orbitals as per Eq. (\ref{proj1-1D-f}). Right: Continuous Fourier modes on a ring corresponding to Eq. (\ref{proj1-1D-f}), for ${\cal G}^6$, labeled by representation index $j$. Degenerate pairs arise from modes $j$ and $−j$, while the $j=3$ mode is self-conjugate and non-degenerate. Sampling the continuous modes at atomic positions yields the symmetry-adapted linear combinations shown on left. Each molecular orbital on the left corresponds directly to a Fourier mode with the same index $j$ on the right. }
    \label{fig:benzene}
\end{figure*}
To make the correspondence between the group-theoretic formalism and physical wavefunctions explicit, we present in Fig. \ref{fig:benzene}, a side-by-side comparison of molecular orbitals and their underlying Fourier structure. The left figure in Fig. \ref{fig:benzene} shows the $\pi$ molecular orbitals of benzene, expressed as symmetry-adapted linear combinations (SALCs) of atomic orbitals and labeled. On the right, we show the corresponding continuous Fourier modes $\exp{\imath j x}$ on a ring, labeled by the same index $j$ appearing in Eq. (\ref{proj1-1D-f}). The nodal structure and phase relationships in the molecular orbitals map directly onto the Fourier modes, with degenerate orbital pairs corresponding to modes $+j$ and $−j$. In this sense, the molecular orbitals may be viewed as discrete samples of the continuous Fourier basis at the atomic positions, providing a concrete physical realization of the representation-theoretic structure developed in Sections \ref{sec:one} and \ref{sec:two}.

\begin{figure*}
    \centering
    \includegraphics[width=\linewidth]{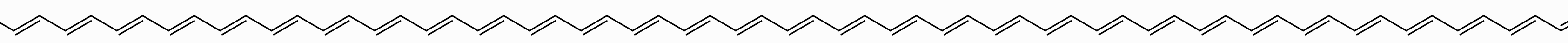}
    \caption{A polyene with ${\cal G}^{15,2}$ symmetry.}
    \label{fig:polyene}
\end{figure*}
We next illustrate this powerful idea above by using a simple linear polyene molecular system. The structure to be inspected is shown in Figure \ref{fig:polyene}.  
If we consider a pair of double bonds (-CH=CH-CH=CH-) to be mapped to a single unit, we may infer that the molecule has 15 repeat units and is periodic beyond the 15 units shown. Furthermore, we wish to now map the molecular unit given by a pair of double bonds (-CH=CH-CH=CH-) to an integer. Thus the entire molecule is now a physical representation of the group ${\cal G}^{15}$. Assuming that the electrons in this system are in a periodic one-dimensional potential with periodicity that extends out to the residues shown in the figure, we may write the overall electronic wavefunction for the system as being composed of single particle wavefunctions\cite{cotton} or molecular orbitals\cite{Albright2013} constructed from symmetry adapted linear combinations (SALCs)\cite{cotton,Albright2013} of atomic orbitals and given by
\begin{align}
f_{\text{SALC}}^{[j]}(x) &= {\cal P}^{(j)} f_{\text{AO}}(x) \nonumber \\
&= \frac{1}{15} \sum_{k=1}^{30}
\exp{\imath (2\pi/15)jk} f_{\text{AO}}(C_{15}^{15-k} x) 
\label{proj1-1D-f-SALC15}
\end{align}
where the functions $f(x)$ are now orbitals corresponding to the individual double bond units (-CH=CH-CH=CH-), and $f(C_{15}^{15-k} x) $ translates the functions to the next unit. The prefactor $\exp{\imath (2\pi/15)jk}$ in turn introduces a phase while constructing the linear combination in Eq. (\ref{proj1-1D-f-SALC15}). 

However, based on Eqs. (\ref{CRT-15-1}) and (\ref{CRT-15-2}) the group, ${\cal G}^{15}$ can be written as direct product of the prime-factor subgroups ${\cal G}^{5}$ and ${\cal G}^{3}$ and as per Eq. (\ref{proj1-1D-f-directproduct}), the projection operators of ${\cal G}^{15}$ can also be written as direct product of the prime-factor subgroup projection operators. That is, 
\begin{align}
f_{\text{SALC}}^{[j]}(x) =& 
\left[ \frac{1}{5} \sum_{k=1}^{5}
\exp{\imath (2\pi/5)jk} P_{C_5^{5-k}} \right] 
\nonumber \\ & 
\left[ \frac{1}{3} \sum_{k=1}^{3}
\exp{\imath (2\pi/3)jk} P_{C_3^{3-k}} \right]
f_{\text{AO}}( x) \nonumber \\ =& 
\frac{1}{15} \sum_{k=1}^{5} \sum_{k'=1}^{3}
\exp{\imath (2\pi/15)(3jk+5jk')} 
\nonumber \\ & P_{C_5^{5-k}} P_{C_3^{3-k'}} 
f_{\text{AO}}( x)
\label{proj1-1D-f-directproduct-C15}
\end{align}
One could also write Eq. (\ref{proj1-1D-f-directproduct-C15}) as
\begin{align}
f_{\text{SALC}}^{[j]}(x) =& 
\frac{1}{15} \sum_{k=1}^{5} \sum_{k'=1}^{3}
\exp{\imath (2\pi/15)(3jk+5jk')} 
\nonumber \\ & 
f_{\text{AO}}( C_5^k C_3^{k'}x)
\label{proj1-1D-f-directproduct-1}
\end{align}
But proceeding further, we have, 
\begin{align}
f_{\text{SALC}}^{[j]}(x) =& 
\frac{1}{15} \sum_{k=1}^{5} \sum_{k'=1}^{3}
\exp{\imath (2\pi/15)j(3k+5k')} 
\nonumber \\ & P_{C_{15}^{15-3k}} P_{C_{15}^{15-5k'}} 
f_{\text{AO}}( x)
\nonumber \\ =& 
\frac{1}{15} \sum_{k=1}^{5} \sum_{k'=1}^{3}
\exp{\imath (2\pi/15)j(3k+5k')} 
\nonumber \\ & P_{C_{15}^{30-(3k+5k')}} 
f_{\text{AO}}( x)
\label{proj1-1D-f-directproduct-2}
\end{align}
As can be seen from Eq. (\ref{CRT-15-2}), the product of operations in ${\cal G}^5$ and ${\cal G}^3$ yield individual operations in ${\cal G}^{15}$. As a result, all pairs of operations, $P_{C_{15}^{15-3k}} P_{C_{15}^{15-5k'}} \equiv P_{C_{15}^{30-(3k+5k')}}$. where $P_{C_{15}^{15-3k}}$ belongs to ${\cal G}^5$ and $P_{C_{15}^{15-5k'}}$ belongs to ${\cal G}^3$ together yield a single operation in ${\cal G}^{15}$. Thus the right side of Eq. (\ref{proj1-1D-f-directproduct-C15}) is essentially identical to the right side of Eq. (\ref{proj1-1D-f-SALC15}) but now generated using the prime factor sub-groups. Alternately, if we make the substitution, $(3k+5k') = k''$, then it follows that
\begin{align}
f_{\text{SALC}}^{[j]}(x) =& 
\frac{1}{15} \sum_{k''=1}^{15} 
\exp{\imath (2\pi/15)jk''} 
\nonumber \\ & P_{C_{15}^{15-k''}} 
f_{\text{AO}}( x) \nonumber \\
=& 
\frac{1}{15} \sum_{k''=1}^{15} 
\exp{\imath (2\pi/15)jk''}  
f_{\text{AO}}(C_N^{k''} x)
\label{proj1-1D-f-directproduct-3}
\end{align}
and hence the prime factor subgroups completely capture the symmetry of the functions. 

\subsection{Connecting back to Shor's algorithm}
Now, after all the polyene we constructed above has 30 double-bonded units and so, we could equally well have interpreted the periodicity as being part of the group ${\cal G}^{N,a} \equiv {\cal G}^{15,2}$. Then, to connect to the discussion of the hidden subgroup problems in the sections above, one may introduce cosets using Figure \ref{fig:Coset-G-15}. The requirement that all rotations inside a coset should have the same function value presents additional symmetry constraints on top of the Bloch function requirement above. That is, there are specific irreducible representations within the polyene that will have similar symmetry properties as the Eqs. (\ref{HSB1})-(\ref{HSB4}). To be precise, from some
\begin{align}
x_1 \rightarrow (i_1,j_1) \equiv C_{30}^{i_1}  C_{2}^{j_1} \rightarrow f_{\text{SALC}}^{[j]}(x_1)
\label{HSB-SALC-1}
\end{align}
and 
\begin{align}
x_2 \rightarrow (i_2,j_2) \equiv C_{30}^{i_2}  C_{2}^{j_2} \rightarrow f_{\text{SALC}}^{[j]}(x_2)
\label{HSB-SALC-2}
\end{align}
\begin{align}
f_{\text{SALC}}^{[j]}(x_1) = f_{\text{SALC}}^{[j]}(x_2)
\label{HSB-SALC}
\end{align}
Thus implies that both $x_1$ and $x_2$ map to the same SALC value which is the essence of Eqs. (\ref{HSB1})-(\ref{HSB4}). Based on such measurements one may be able to map out the irreducible representation elements described in the previous section and thus arrive at the prime factors based on constructions that exploit such symmetry. 




\section{Discussion and Outlook}
\label{sec:seven}
We utilize the language of group theory to describe Shor's algorithm and relate such a description to symmetries in molecular, and condensed phase  systems. By mapping a truncated set of integers to a unit circle, we show that the prime factorization process becomes similar to finding the fundamental and basic rotation operations that leave functions defined on the circle to be invariant. Here we utilize the well known Chinese Remainder theorem in mathematics where it is shown that the powers of prime factor sub-groups of $\mathbb{Z}_N$ can be used reconstruct the group and hence irreducible representations of the prime factor sub-groups of $\mathbb{Z}_N$ are contained within functions that transform according to $\mathbb{Z}_N$. We show that identical properties exist for cyclic (Abelian) groups that then may be mapped to symmetries in molecules and condensed phase assemblies. We also show how the {\em Oracle} in Shor's algorithm which returns the function value $f(x) = a^x \pmod{N}$ is one example of such a function. Using Bloch's theorem from molecular and condensed matter physics, we show that molecular orbitals and wavefunctions in condensed phase assemblies obey similar symmetry properties (as the oracle in SHor's algorithm), thus opening the possibility that synthetic systems may be tailored and designed such that these {\em contain} within them solutions to hard mathematical problems, such as prime-factoring. We also propose introduce molecules that may have such properties. Given the broad significance of prime-factory in areas such as encryption, cyber-security, communication and finance, we expect that our work will be impactful through the fundamental connections thus exposed.

\section{Acknowledgement}
This research was supported by the National Science
Foundation grants CHE-2102610 to SSI. The authors acknowledge Mr. Ata Tuncer, Prof. Babak Seradjeh, and Prof. Chen-ting Liao for their constructive comments on this work.  

\input{refs.text}
\end{document}

%% file: Figures/Shor-ckt.tex
\begin{quantikz}[slice all,row sep=0.7cm,column sep=1cm]
   \lstick{$\ket{0}^{\otimes q}$} &[3em] 
   \gate{H^{\otimes q}}\qwbundle{q} & 
   \gate[wires=2][2cm]{U_f}\gateinput{$x$}\gateoutput{$x$} &
   \qw &
   \gate{\mathit{QFT}} &
   \meter{} & 
   \rstick{$\tilde{v}$}\cw
   \\
   \lstick{$\ket{0}^{\otimes m}$} &[3em]
   \qw\qwbundle{m} &
   \gateoutput{$a^x \pmod{N}$} &
   \meter{} &
   \rstick{$w$}\cw 
\end{quantikz}